\documentclass[preprint]{aastex}

\newcommand{\kms}{km~s$^{-1}$}
\newcommand{\ms}{m~s$^{-1}$}
\newcommand{\degrees}{$^{\circ}$}
\newcommand{\omorb}{\ensuremath{\omega_{\mathrm{orb}}}}
\newcommand{\omrot}{\ensuremath{\omega_{\mathrm{rot}}}}
\newcommand{\Porb}{\ensuremath{P_{\mathrm{orb}}}}
\newcommand{\Prot}{\ensuremath{P_{\mathrm{rot}}}}
\newcommand{\tss}{\ensuremath{\theta_{\mathrm{ss}}}}
\newcommand{\HDo}{HD~189733b}
\newcommand{\HDt}{HD~209458b}

\begin{document}

\title{The Atmospheric Circulation and Observable Properties \\
	of Non-Synchronously Rotating Hot Jupiters}

\author{Emily Rauscher
  \\ \textit{Department of Astrophysical Sciences, Princeton University
  \\ 4 Ivy Lane, Peyton Hall, Princeton, NJ 08544, USA}
  \\ and
  \\ Eliza M. R. Kempton
  \\ \textit{Department of Physics, Grinnell College
  \\ Noyce Science Building, Grinnell, IA 50112, USA}}
  
\begin{abstract}

We study the feasibility of observationally constraining the rotation rate of hot Jupiters, planets that are typically assumed to have been tidally locked into synchronous rotation.  We use a three-dimensional General Circulation Model to solve for the atmospheric structure of two hot Jupiters (\HDo~and \HDt), assuming rotation periods that are 0.5, 1, or 2 times their orbital periods (2.2 and 3.3 days, respectively), including the effect of variable stellar heating.  We compare two observable properties: 1) the spatial variation of flux emitted by the planet, measurable in orbital phase curves, and 2) the net Doppler shift in transmission spectra of the atmosphere, which is tantalizingly close to being measurable in high-resolution transit spectra.  Although we find little difference between the observable properties of the synchronous and non-synchronous models of \HDo, we see significant differences when we compare the models of \HDt.  In particular, the slowly rotating model of \HDt~has an atmospheric circulation pattern characterized by westward flow and an orbital phase curve that peaks after secondary eclipse (in contrast to all of our other models), while the quickly rotating model has a net Doppler shift that is more strongly blue-shifted than the other models.  Our results demonstrate that the combined use of these two techniques may be a fruitful way to constrain the rotation rate of some planets, and motivate future work on this topic.

\end{abstract}

\section{Introduction}

Perhaps the most commonly used assumption in studying hot Jupiters is this: that the planets with zero eccentricity have been tidally locked into synchronous rotation, meaning that their orbital and rotational periods are equal, with one hemisphere in unending day and the other in perpetual night.  This assumption is based on timescale arguments that tidal forces should lock a planet into a state of synchronous rotation long before they circularize its orbit \citep[e.g.,][]{Rasio1996}.  However, the physical mechanism of tidal dissipation within a gas giant is a complex process and an area of active research \citep[e.g.,][]{Ogilvie2004,Remus2012,Weinberg2012}.  Recently \citet{Socrates2012} have argued that the simplifying assumptions commonly used to model the tidal response of hot Jupiters may not accurately capture the physics at work, especially when it comes to the tidal circularization of highly eccentric gas giants.  Additionally, even if a planet were tidally locked into a synchronous state, the flow of angular momentum between the planet's orbit, atmosphere, and/or interior could torque the planet out of synchronous rotation \citep{Showman2002}.  In particular, \citet{Arras2010} have proposed that the ``thermal tide'' torque originally used to explain Venus' spin state \citep{Gold1969} could be at work in hot Jupiters, resulting in a steady state with asynchronous rotation and a source of internal heating through the constant dissipation of gravitational tides.  Since many hot Jupiter radii are observed to be larger than can be explained without an extra source of heating inside the planet,\footnote{\citet{Spiegel2013} contains a good, recent description of this open question and compares the proposed solutions.} it is especially important to question whether there are any observable signatures that could be used to identify non-synchronous rotation of hot Jupiters.

A gas giant planet locked into synchronous rotation can still support differential fluid motions throughout its atmosphere and interior; the synchronicity is a requirement on the total angular momentum of the body.  If a solid core exists within a synchronously rotating planet, it will presumably be locked into solid-body rotation at the synchronous rate and dominate the (non-orbital) angular momentum of the planet.  The core is in continuous contact with the fluid layers above and angular momentum transport will ensure that motion in the upper atmospheric layers is influenced by the global rotation of the planet.  The atmospheric patterns of winds and (observable) temperature structures are strongly dependent on the rotation rate, through the influence of the Coriolis force on atmospheric dynamics.

Given the assumption of synchronous rotation, the Coriolis force should have a much weaker effect on the atmospheres of hot Jupiters ($\Prot \sim 3$ days) than on the atmosphere of the more quickly rotating namesake, Jupiter ($\Prot \sim 10$ hours), and early estimates of dynamical quantities predicted that this should result in very large atmospheric features on hot Jupiters \citep{Showman2002,Cho2003,Menou2003}.  This has been born out in three-dimensional General Circulation Models (GCMs) of the atmospheres of synchronously rotating hot Jupiters \citep{Showman2009,Thrastarson2011,Perna2012,Rauscher2012b,DobbsDixon2013,Mayne2014}.  One common atmospheric characteristic found in most GCMs is a super-rotating\footnote{``Super-rotating'' means that the flow is eastward, with greater angular momentum than the planet's rotation at that radius and latitude.} equatorial jet, which results in the hottest region of the atmosphere being advected to the east of the substellar point.  The development of the eastward jet is a general property of the hemispheric forcing regime of hot Jupiters \citep{Showman2011}, although it can be inhibited by the presence of high viscosity or strong drag \citep{DobbsDixon2010,Perna2010a,Showman2011,Rauscher2013}, in which case the hottest region remains near the substellar point.  The lack of a shift from the substellar point does not necessarily preclude the existence of an equatorial jet, however, since in the case of very short radiative timescales (i.e., for very hot atmospheres), the gas at the substellar point cools more quickly than winds can advect it away from the substellar point \citep{Perna2012,Showman2013}.

For a few of the brightest hot Jupiters it has been possible to measure the flux emitted from the planet as a function of longitude, by observing the planetary system throughout an orbit and watching for the varying flux as different phases come into view.  Of the measured orbital phase curves for hot Jupiters on circular orbits, we see the expected range of behavior:\footnote{Note that the westward phase shift seen in the orbital phase curve for the hot Jupiter Kepler-7b was observed at optical wavelengths and is interpreted as reflected light from spatially inhomogeneous clouds \citep{Demory2013}, which is not inconsistent with the presence of an eastward equatorial jet.} some planets have their brightest regions well aligned with the substellar point \citep{Cowan2007,Snellen2009,Welsh2010}, while others have the brightest region shifted to the east of the substellar point \citep{Cowan2012,Knutson2012}, including the curious case of $\upsilon$~Andromedae~b \citep{Crossfield2010}, which has a surprisingly large shift ($\sim$80\degrees).\footnote{Naively we would expect the amplitude of the flux variation to be anti-correlated with the amount by which the hot spot is shifted, since the shift is a result of winds transporting the hot gas faster than it can cool, but such efficient winds should result in a more homogenous horizontal temperature structure \citep[e.g.][]{Cowan2011a}.  No current GCM can reproduce the $\upsilon$ And b phase curve.}  The planet with the best signal-to-noise observations, \HDo, also has eclipse mapping measurements that independently support the eastward shift of the brightest region \citep{Majeau2012}, although see \citet{deWit2012} for a careful discussion of parameter degeneracies.

\citet{Showman2009} performed the first simulations of a non-synchronously rotating hot Jupiter, modeling the planet \HDo.  Although there was some variation in the circulation pattern for different rotation rates---which we will return to discuss in detail below---all models developed an eastward jet at the equator and produced orbital phase curves in which the brightest region was shifted east of the substellar point, although the magnitude of the shift varied with the rotation rate \citep[at a level comparable to the shift in the peak emission between models run at different metallicities,][]{Showman2009}.  Their results imply that an orbital phase curve measurement cannot conclusively differentiate between synchronous and non-synchronous rotation.

Finding ways to measure the rotation rate of solar system giant planets has been a challenge in its own right.  The most widely accepted values come from \emph{in situ} satellite measurements of the periodicity of the planetary magnetic field (assumed to be generated in the deep interior and therefore tied to that bulk rotation rate) or modulation of radio emission, presumed to be tied to the rotating magnetic field lines \citep[e.g.,][]{Desch1981,Desch1986,Warwick1989}.  However, additional types of measurements have also been used to estimate the rotation rates, such as: the oblateness of a planet, Doppler-broadening of line widths in atmospheric spectra, periodicity in light curves, and tracking of imaged cloud features \citep[see a discussion in the review by][]{Stevenson1982}.  The rotation and global structure of the solar system giant planets continues to be an area of active research; one example being the recent work by \citet{Helled2010}, which calls into question whether the magnetic rotation periods are tied to the bulk rotation rates of Uranus and Neptune, since those periods do not match the ones derived from fitting to the planets' oblateness measurements.

It may be a long wait before we can achieve \emph{in situ} measurements of exoplanet properties, but the other methods of measuring planetary rotation rates may be possible in the nearer future.\footnote{The first measurement of an exoplanet rotation rate was published during the refereeing process for this paper.  \citet{Snellen2014} observed rotational broadening in their high-resolution spectrum of the directly imaged young exoplanet $\beta$~Pictoris~b, corresponding to an equatorial rotational velocity of 25$\pm$3 \kms.}  There is a long history of searching for radio emission from exoplanets, which has so far resulted only in upper limits \citep[aside from a very recent possible first detection,][]{Lecavelier2013}.  In order to use radio emission to estimate rotation rates a modulated signal is needed, which may only be detectable with current instruments for strong planetary magnetic field strengths and fortuitous viewing geometry \citep{Hallinan2013}.  Even if a periodic signal were detected, it might be difficult to confidently identify the period as the rotation period of the planet, rather than some stellar period or a period related to star-planet magnetic interaction.

Rotational periods could potentially be measured for Earth-like exoplanets by observing variation in the reflected light from the planet with next-generation instruments \citep{Palle2008}, but for close-in planets the periodic variation in thermal emission is subject to several parameters, including the radiative timescale of the atmosphere and wind speeds on the planet \citep{Cowan2011a}.  For hot Jupiters we expect the period of variation to be well matched to the orbital period, regardless of the rotation rate \citep[based on the work of][discussed above]{Showman2009}.  The oblateness of a planet could possibly be measured by the detailed shape of the light curve as the planet transits its star \citep{Seager2002,Barnes2003}; however, a recent observational limit could only constrain the rotation period of the hot Jupiter HD 189733b to be greater than about 10 hours, or $\sim$20\% of the synchronous rotation period \citep{Carter2010}.

Here we investigate the method of using Doppler shifts in a hot Jupiter's transit spectrum to observationally constrain its rotation rate.  Previous work on this topic has shown that the Doppler rotational broadening in the spectrum may be of comparable magnitude to the expected shifts from atmospheric winds \citep{Brown2001,Spiegel2007,Kempton2012,Showman2013}, meaning that both effects must be included in any predictive model.  In this paper we compare models for hot Jupiters at different rotation rates, consistently including the atmospheric circulation structure calculated from a 3D GCM using that rotation rate.  In Section~\ref{sec:model} we discuss our modeling framework, both the radiative transfer with Doppler shifts \citep[from][]{Kempton2012} and the GCM \citep[from][]{Rauscher2012b}, now adapted to allow for non-synchronous rotation.  In Section~\ref{sec:circ} we present the results of the GCMs and discuss the variation in circulation patterns between models.  We show the observable properties of each model in Section~\ref{sec:obs}: the thermal phase curves and the transit spectra.  Finally, in Section~\ref{sec:conc} we summarize our results and discuss the potential of observationally constraining hot Jupiter rotation rates.

\section{Model set-up} \label{sec:model}

In this paper we compare synchronous and non-synchronous models of the two best-known hot Jupiters: \HDo~and \HDt.  We use our three-dimensional atmospheric dynamics code with double-gray radiative transfer \citep[for a detailed description see][and references therein]{Rauscher2012b}.  This GCM solves the standard set of inviscid fluid equations as applied to a rotating atmosphere in vertical hydrostatic equilibrium: the primitive equations of meteorology, with pressure as the vertical coordinate.  The horizontal components of these equations are solved in spectral space, finite differencing is used for the vertical components, and hyperdissipation\footnote{Hyperdissipation is a common numerical technique whereby a high-order operator is applied to the temperature and flow fields.  See \citet{Rauscher2012b} for a discussion of how we chose the strength of hyperdissipation to apply to our models, as well as some of the complexities involved.} is used to prevent the build-up of noise on the smallest resolved scales.  The radiative heating is calculated using standard two-stream vertical radiative transfer, with two constant absorption coefficients: one for the optical band and one for the infrared.  At high optical depth the radiative transfer transitions to a flux-limited diffusion scheme, using the infrared opacity.

We continue to use the same physical parameters for \HDt~and \HDo~as we did in our previous, synchronous models of these planets \citep{Rauscher2012b,Rauscher2013}, which are reported in Table~\ref{tab:params}.  Note that we use the same optical and infrared absorption coefficients for both planets, as well as the same internal heat flux.  In lieu of a consensus on the atmospheric compositions of \HDo~and \HDt, we choose to use absorption coefficients that give radiative equilibrium temperature-pressure profiles similar to those calculated from more complex 1D radiative transfer codes.  We found that the circulation and temperature structure of the planet's observable pressure layers is not sensitive to the internal heat flux, over a range of reasonable values \citep{Rauscher2012b}.

For all of the results presented in this paper, the horizontal resolution was chosen to match our previous fiducial models: T31, corresponding to $\sim$4\degrees~in latitude and longitude.  We performed limited runs of the non-synchronous models at higher resolution, T42, but found no significant differences in the atmospheric flow and temperature structures.\footnote{\citet{Polichtchouk2012} argue that higher resolution (at least T85) is required to capture instabilities in hot Jupiter atmospheres, but their analysis is based on models without any diabatic (radiative) forcing and it is not clear how the instabilities and resolution requirements translate to hot Jupiter atmospheres with strong stellar forcing and very short radiative timescales.}  We also used the same vertical resolution, but extended our upper boundary to lower pressures, now with 45 levels logarithmically spaced in pressure, from 100 bar to 10 microbar.  In all cases the simulations were initialized with zero winds and a temperature profile that was uniform horizontally and varied with pressure according to the analytic formalism of \citet{Guillot2010}, with the averaging parameter $f=0.375$, in-between a global average and a dayside-only average.  In all cases we ran the simulation for 1000 \Porb, by which point the lower pressure (observable) layers have reached a statistically steady state.  We know from our previous, longer runs that the deep pressure levels continue to slowly accelerate, but that the flow in the upper layers is not significantly different at 1000 \Porb~from what we see at 2000 \Porb.\footnote{This gradual acceleration at depth is a slow rise in the kinetic energy of those levels, because of increasing wind speeds, but the global momentum is unchanged, with westward flow accelerating to balance the accelerating eastward winds.  Since the deeper levels dominate the mass of the modeled atmosphere, small adjustments at depth are sufficient to balance flow in the upper levels.}
For each planet (\HDo, \HDt) we used the same length time step (40, 62 seconds) and hyperdissipation timescale (950, 1500 seconds) for the three differently rotating models.

\begin{deluxetable}{lccc}
\tablewidth{0pt}
\tablecaption{Model parameters used}
\tablehead{
\colhead{Parameter}  &  \colhead{HD 189733b} & \colhead{HD 209458b}  & \colhead{Units}
}
\startdata
Radius of the planet, $R_p$ 	& $8\times 10^7$ 		& $1\times 10^8$ 		& m \\
Gravitational acceleration, $g$ & 22					& 8 					& m s$^{-2}$ \\
Orbital rotation rate, \omorb 		& $3.3 \times 10^{-5}$ 	& $2.1 \times 10^{-5}$ 	& s$^{-1}$ \\
\ \ \ Corresponding period, \Porb & 2.2 & 3.3 	& day$_{\oplus}$ \\
Incident flux at substellar point, $F_{\downarrow \mathrm{vis}, \mathrm{irr}}$ & $4.74 \times 10^5$ & $1.06 \times 10^6$  & W m$^{-2}$  \\
\ \ \ Corresponding temperature  & 1700 & 2078 & K \\
\hline
Internal heat flux, $F_{\uparrow \mathrm{IR}, \mathrm{int}}$ & \multicolumn{2}{c}{3500} & W m$^{-2}$ \\
\ \ \ Corresponding temperature  & \multicolumn{2}{c}{500} & K \\
Optical absorption coefficient, $\kappa_{\mathrm{vis}}$ & \multicolumn{2}{c}{$4 \times 10^{-3}$} & cm$^2$ g$^{-1}$ \\
Infrared absorption coefficient, $\kappa_{\mathrm{IR},0}$  & \multicolumn{2}{c}{$1 \times 10^{-2}$}  & cm$^2$ g$^{-1}$ \\
Infrared absorption powerlaw index, $\alpha$ & \multicolumn{2}{c}{0} & -- \\
Specific gas constant, $\mathcal{R}$ & \multicolumn{2}{c}{3523} & J kg$^{-1}$ K$^{-1}$ \\
Ratio of gas constant to heat capacity, $\mathcal{R}/c_P$ & \multicolumn{2}{c}{0.286} & -- \\
\enddata
\label{tab:params}
\end{deluxetable}
\clearpage

\subsection{Non-synchronous rotation} \label{sec:nonsync}

\begin{figure}[ht!]
\begin{center}
\includegraphics[width=\textwidth]{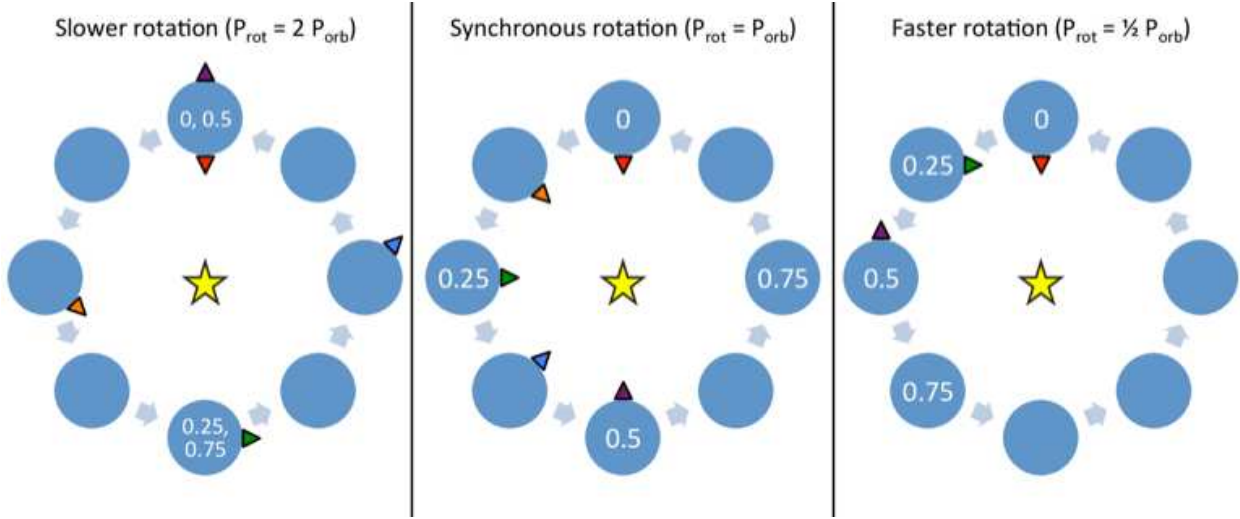}
\end{center}
\caption{Diagram of rotation models.  The northern rotational and orbital poles are pointed out of the page.  In each panel the triangles mark a constant location (longitude) on the planet, with the colors indicating equal fractions of the planet's rotation period.  Similarly, the numbers mark the points at which the planet has completed that fraction of a rotation in an intertial frame.} \label{fig:diagram}
\end{figure}

All GCM models are solved in the frame rotating with the planet (and all wind speeds are quoted within this frame).  The rotation rate of the planet influences the model in two ways: 1) the horizontal momentum equation contains a term for the Coriolis force, and 2) the variation of the irradiation pattern with time is determined by a combination of the rotation and orbital rates.  We assume an obliquity of zero for the planet, meaning that the rotational and orbital planes are aligned and the substellar point is always along the equator.  The longitude of the substellar point as a function of time is given by:
\begin{equation}
\tss = (\omorb - \omrot) t = 2 \pi (\Porb^{-1}-\Prot^{-1}) t. \label{eqn:tss}
\end{equation}
\noindent For a planet locked into a synchronous state, the rotational and orbital periods are equal ($\Prot=\Porb$) and one hemisphere remains continuously irradiated.  If a planet's rotation rate is faster than its orbital rate, the substellar point will move westward (in the direction of decreasing longitude, as is the case for the Earth), while eastward movement results from an orbital rate faster than the rotation.  Figure~\ref{fig:diagram} diagrams systems rotating slower than, equal to, and faster than the orbital rate.  

The stellar flux as a function of latitude ($\phi$), longitude ($\theta$), and pressure ($P$) is given by:
\begin{equation}
F_{\downarrow \mathrm{vis}} (\phi,\theta,P) =  F_{\mathrm{inc}} \cos (\phi) \cos(\theta - \tss) \exp\left( - \frac{1}{\cos (\phi) \cos(\theta - \tss)} \frac{\kappa_{\mathrm{vis}}}{g} P\right)
\end{equation}
\noindent on the day side, where $\cos (\phi) \cos(\theta - \tss) > 0$, and equal to zero on the night side.  Assuming zero albedo, $F_{\mathrm{inc}}$ is the flux incident on the top of the atmosphere at the substellar point.  This is a commonly used form for the absorption of optical light in a non-scattering atmosphere with a well-mixed absorber \citep[e.g.,][]{Stephens1984,Rauscher2012b}.  The cosine of the zenith angle appears twice, once outside of the exponential to account for the geometry of the irradiation (surface areas near the terminator intercept less starlight), and once inside of the exponential to recreate the effect of a longer optical path, since in our modeling scheme the flux in each column of the atmosphere is calculated solely in the vertical direction.

In this paper we present models of \HDo~and \HDt~that are off from synchronous rotation by a factor of 2; we ran models for each planet with $\Prot = 0.5$, 1, and 2 \Porb.  These values were not chosen because we have any reason to expect that gas giants should be tidally locked into higher-order spin-orbit resonances, but because once we put aside the expectation that hot Jupiters should be locked into synchronous rotation, there is no strong theoretical guidance as to what rotation rates might be expected.  As such, we chose to keep the rotation rate within a factor of 2 from synchronous and match rotation states previously modeled by \citet{Showman2009} in order to facilitate comparison with their models.

\subsection{Calculation of observable properties}

One product of our simulation is a self-consistent map of the flux emitted from the top of the atmosphere, as a function of latitude and longitude around the planet.  Due to our modeling assumptions, all of the flux emitted by the planet is in a single, infrared band.  It is trivial to calculate the orbital phase curve that would be measured, by integrating over the hemisphere facing the observer for snapshots from each simulation, throughout one orbit of the planet \citep[for a more detailed description of the radiative transfer see][]{Rauscher2012b}.  

For the transit spectra see \citet{Kempton2012} for details, but the short of it is as follows.  The transmission spectrum arises when the light of the host star passes through the optically thin upper layers of an exoplanet atmosphere as the planet transits.  The excess absorption resulting from this process produces a spectral fingerprint of the planetary atmosphere on the star light obtained during transit.  Because the stellar light follows an oblique trajectory through the planetary atmosphere (rather than a radial path), the geometry of the light rays must be carefully accounted for.  For the models described in this paper, we intercept 5,760 individual light rays through the annulus of the planet's atmosphere (60 concentric circles of 96 rays each, spanning the annulus of the observable portion of the planet's atmosphere).  We calculate the attenuation of stellar light along each ray by a straightforward integration of the radiative transfer equation for the case of absorption only.  We then integrate over the solid angle subtended by each grid cell (represented by a single ray of light) to determine the net absorption produced by the planetary atmosphere as a whole -- the transit spectrum.  For the 3D models presented in this paper, we  account for the fact that each individual light ray encounters multiple (radial) temperature-pressure profiles from the GCM along its path through the atmosphere.  The gas opacities are calculated to be self-consistent with the local temperature and pressure at each location in the atmosphere that the ray encounters.  Additionally, the gas opacities are Doppler shifted according to the local line-of-sight motion that results from the combination of the local wind speed (with zonal and meridional components) and the rotation of the planet.

\section{Comparison of circulation patterns} \label{sec:circ}

\begin{figure}[ht!]
\begin{center}
\includegraphics[width=0.32\textwidth]{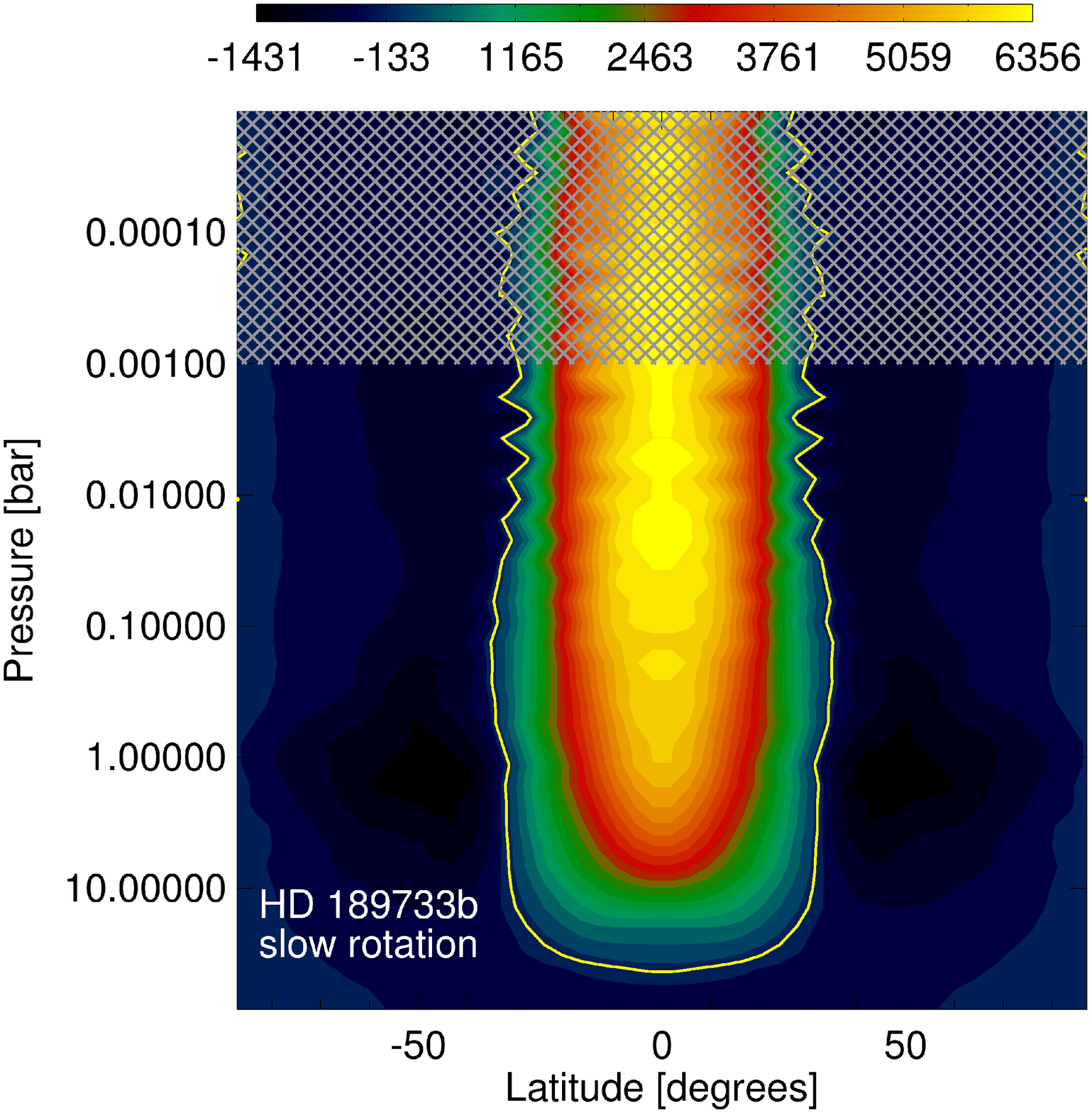}
\includegraphics[width=0.32\textwidth]{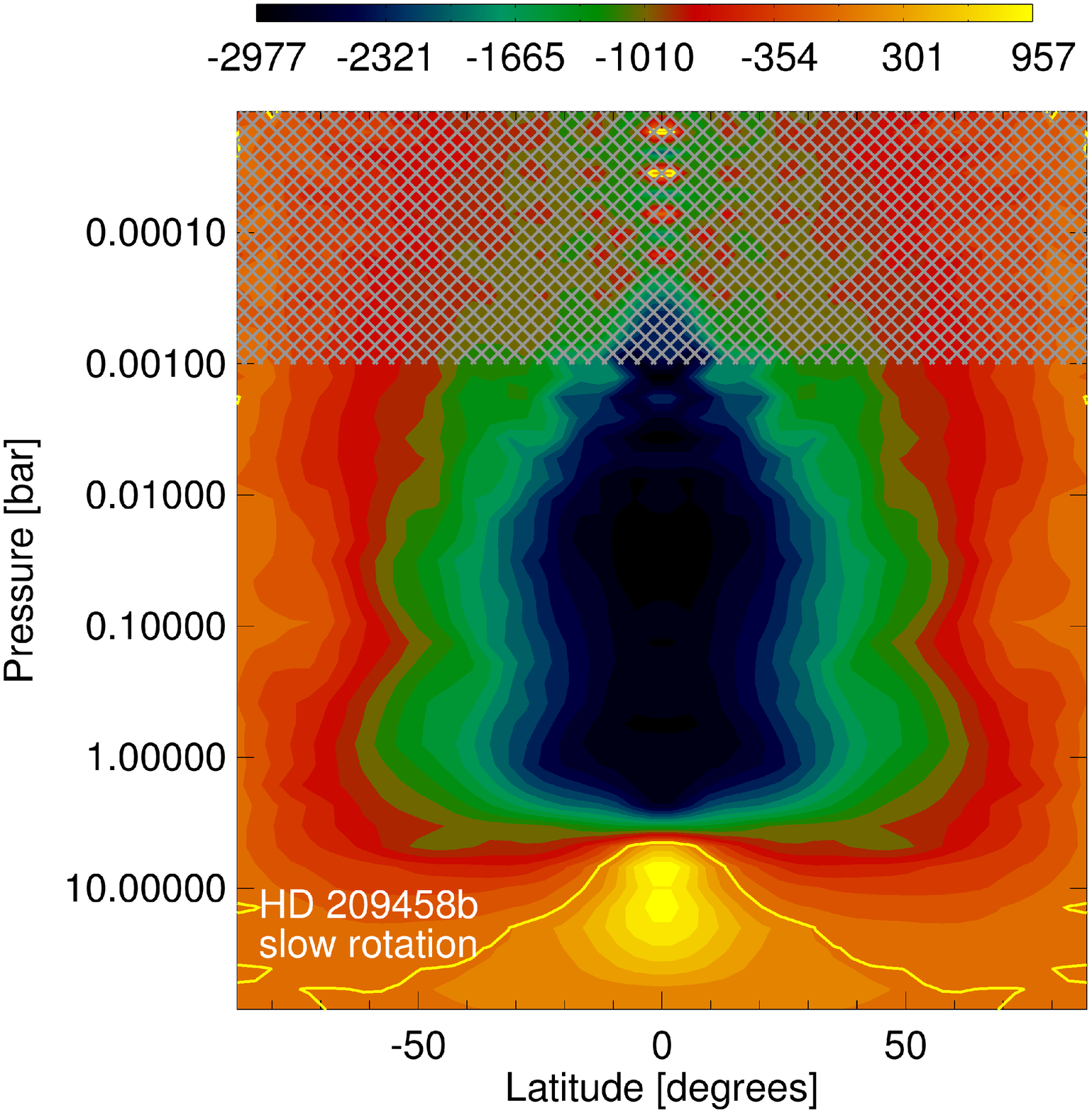} \\
\includegraphics[width=0.32\textwidth]{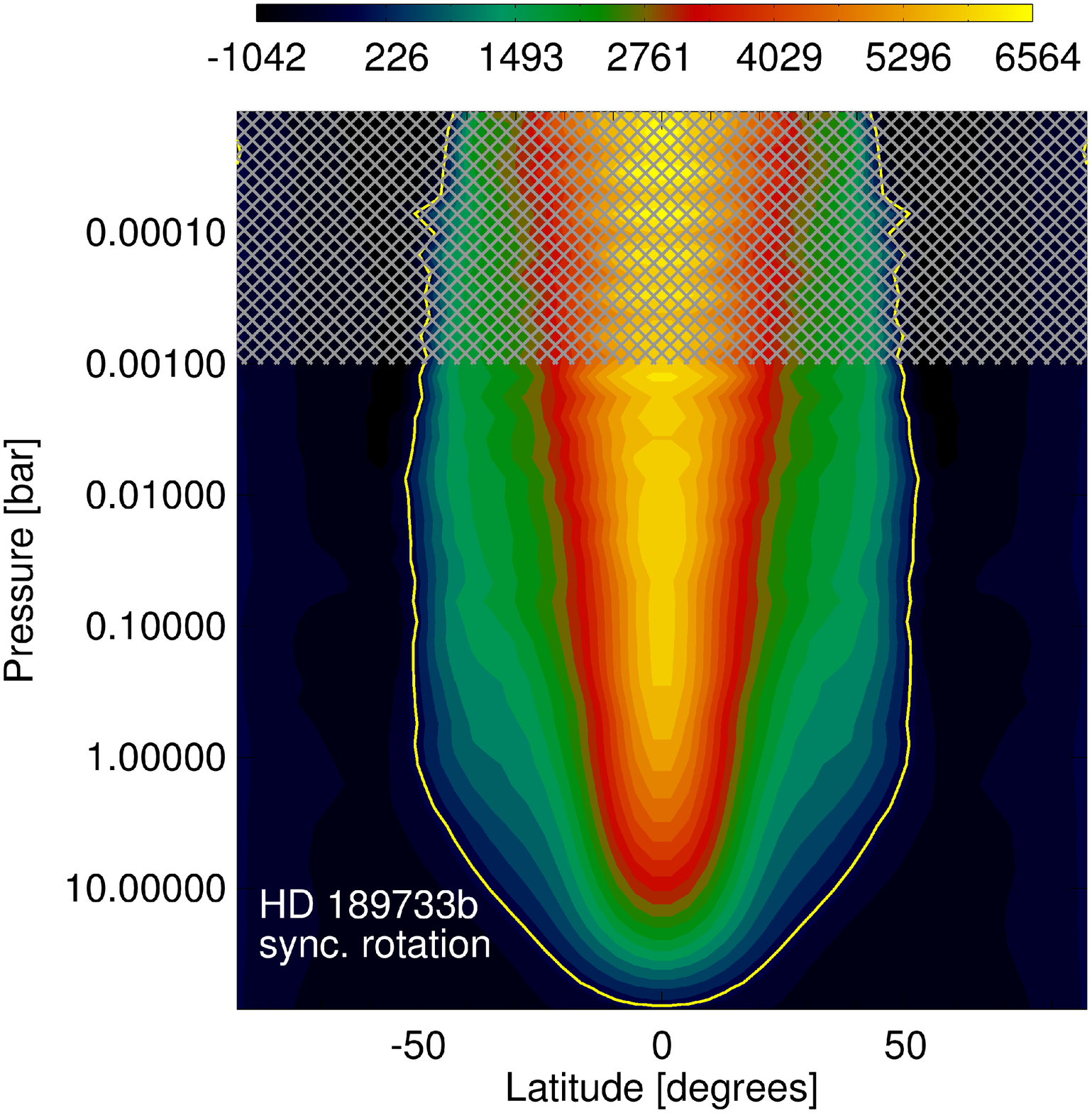}
\includegraphics[width=0.32\textwidth]{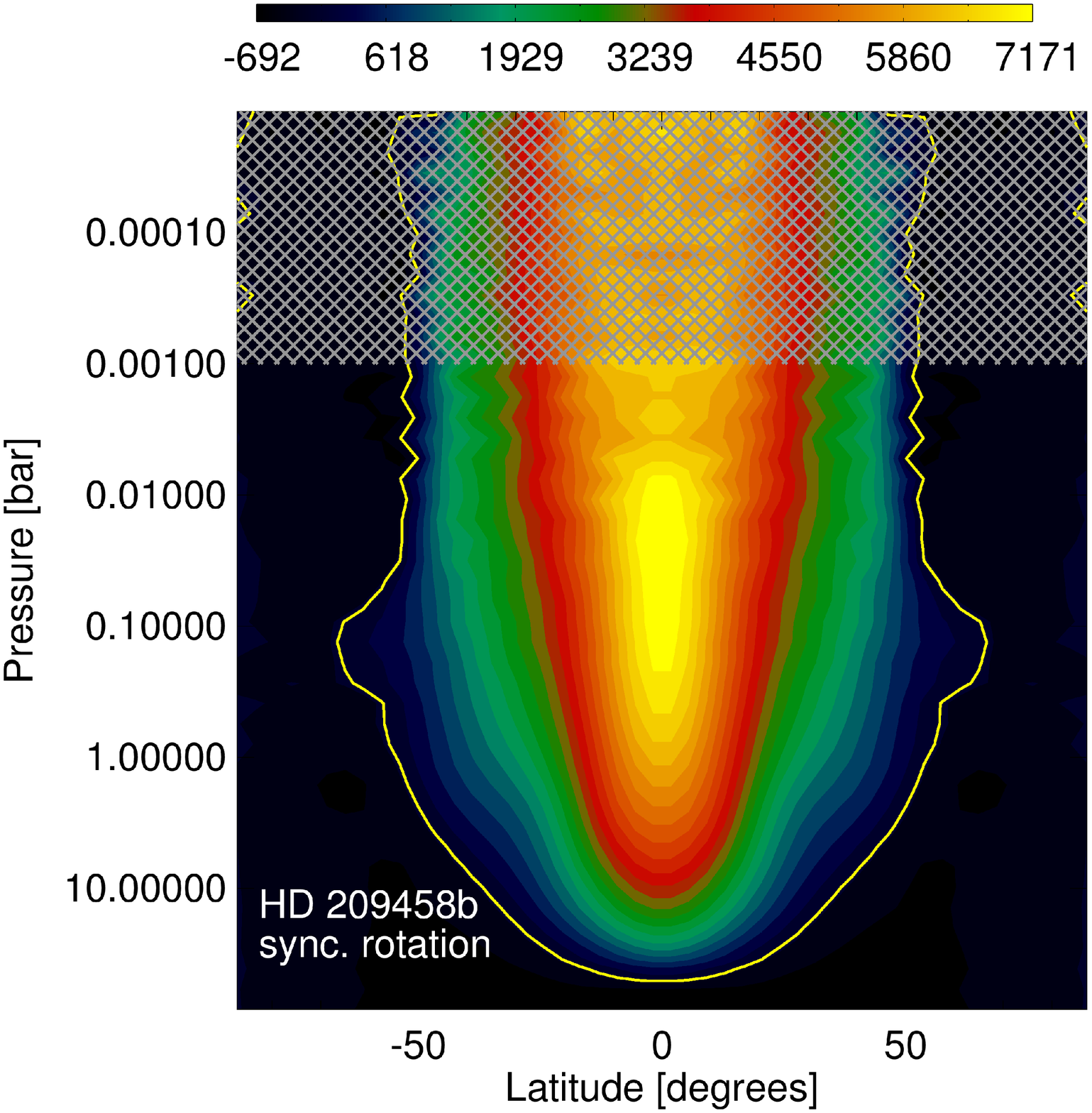} \\
\includegraphics[width=0.32\textwidth]{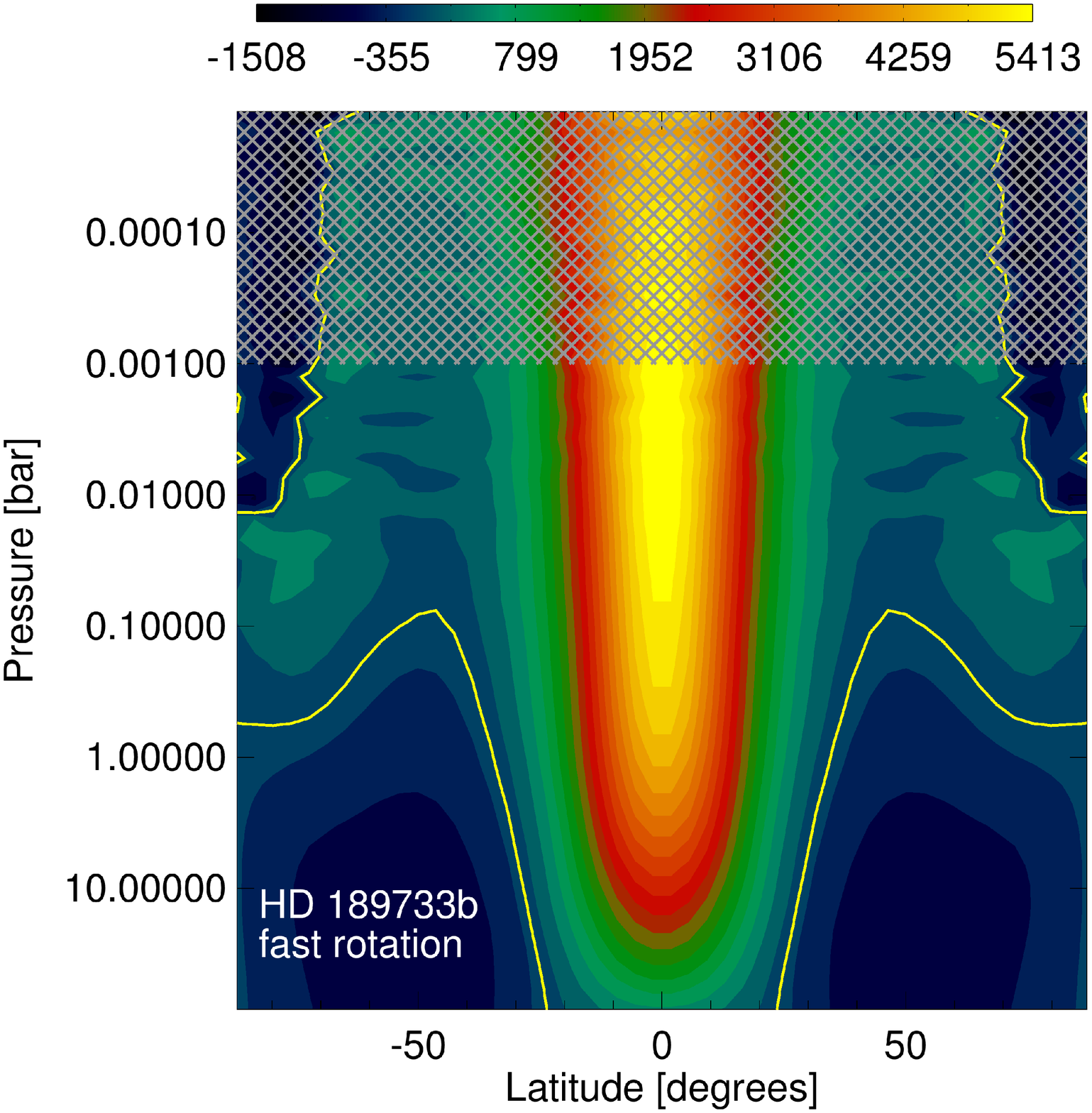}
\includegraphics[width=0.32\textwidth]{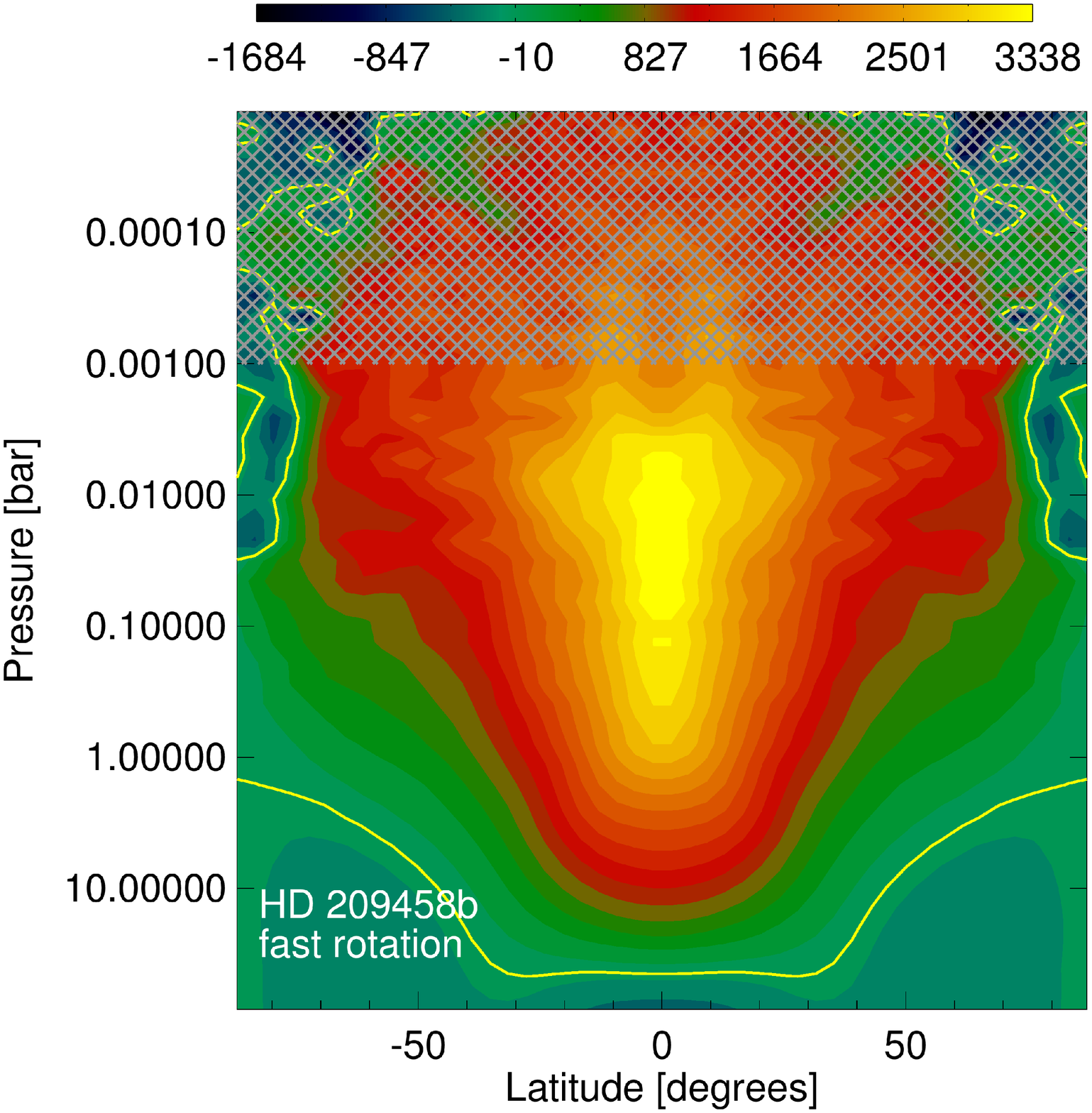}
\end{center}
\caption{Plots of the zonal (east-west) average of the zonal wind (in \ms), as a function of latitude and pressure, for models of \HDo~(\emph{left column}) and \HDt~(\emph{right column}) using (\emph{from top to bottom}) slow, synchronous, or fast rotation rates.  The yellow line separates positive (eastward) from negative (westward) flow.  The gray shading at low pressures is to caution the reader that zonal averages become less informative higher in the atmosphere, as substellar-to-antistellar flow becomes more dominant (so that the winds across each terminator cancel each other in the average).}\label{fig:uz}
\end{figure}

As far as we are aware, \citet{Showman2009} is the only published work that includes GCMs of non-synchronously rotating hot Jupiters on circular orbits \citep[models of eccentric, pseudo-synchronously rotating planets can be found in][]{Lewis2010,Kataria2013}.  Our results for \HDo~are in qualitative agreement with the models for this planet in \citet{Showman2009}.  In all cases the circulation develops the ubiquitous eastward equatorial jet, although the width of the jet is wider in the more slowly rotating models and a second set of high latitude jets forms in the more quickly rotating models \citep[compare our Figure~\ref{fig:uz} with Figures 5 and 14 in][]{Showman2009}.  This is exactly as should be expected, from the role of the Coriolis force in constraining dynamical scales in the atmosphere \citep[see the review by][]{SCM2010}; faster rotation means more, narrower jets.  The synchronous rotation rate of \HDo~is faster than that for \HDt~($\Porb = 2.2$ days, compared to 3.3 days) and so our quickly rotating model of \HDt~is at an intermediate point between the locked and fast models of \HDo, with hints of the additional high-latitude jets that are more clearly apparent in the fast model of \HDo.  Similarly, the equatorial jet in the locked model of \HDt~is slightly wider than the jet in the locked model of \HDo.  However, our slowly rotating model of \HDt~shows a significant circulation regime shift: the flow throughout most of the atmosphere is predominantly westward.

\begin{figure}[ht!]
\begin{center}
\includegraphics[width=0.475\textwidth, trim= 0 0 0 60, clip=true]{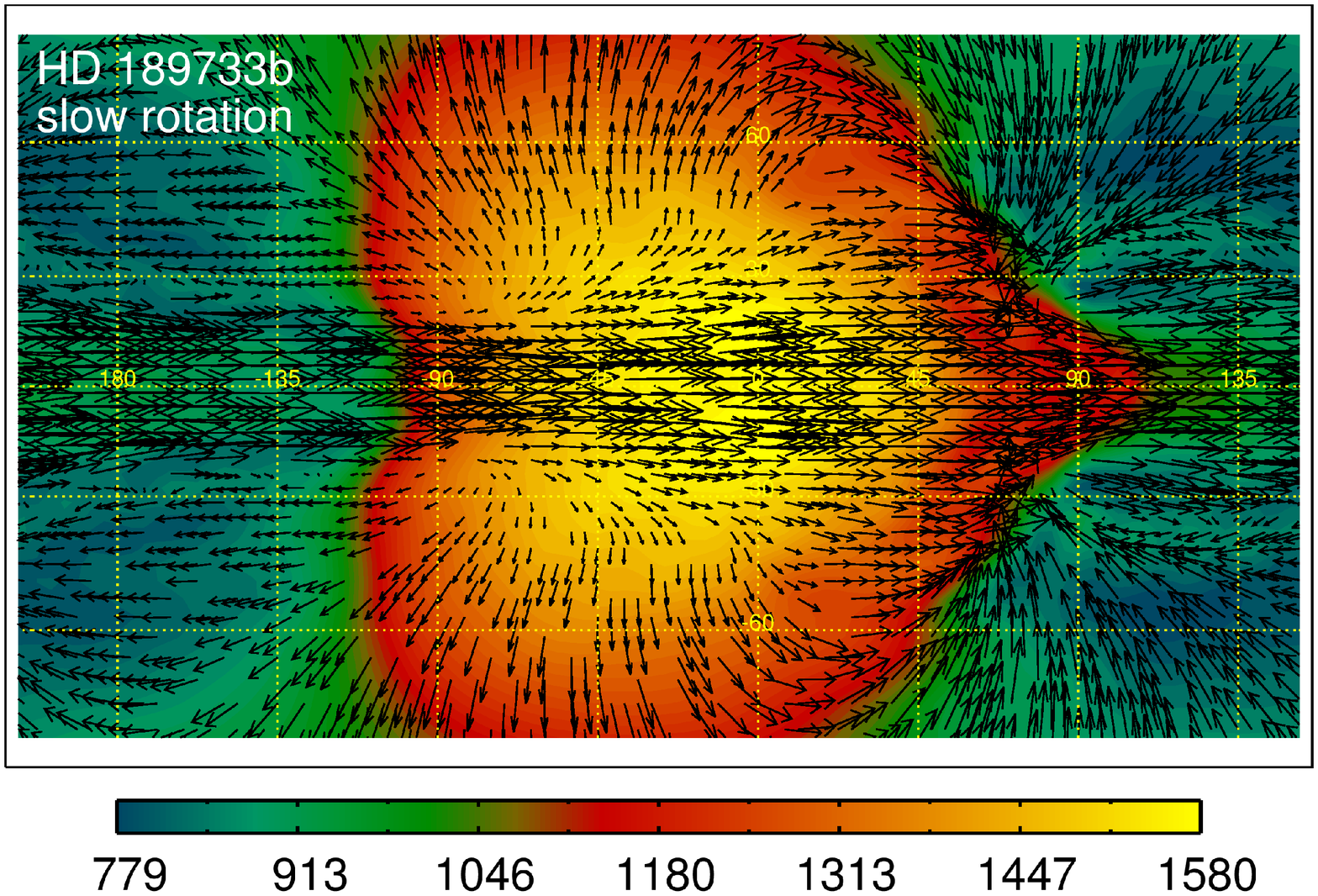}
\includegraphics[width=0.475\textwidth, trim= 0 0 0 60, clip=true]{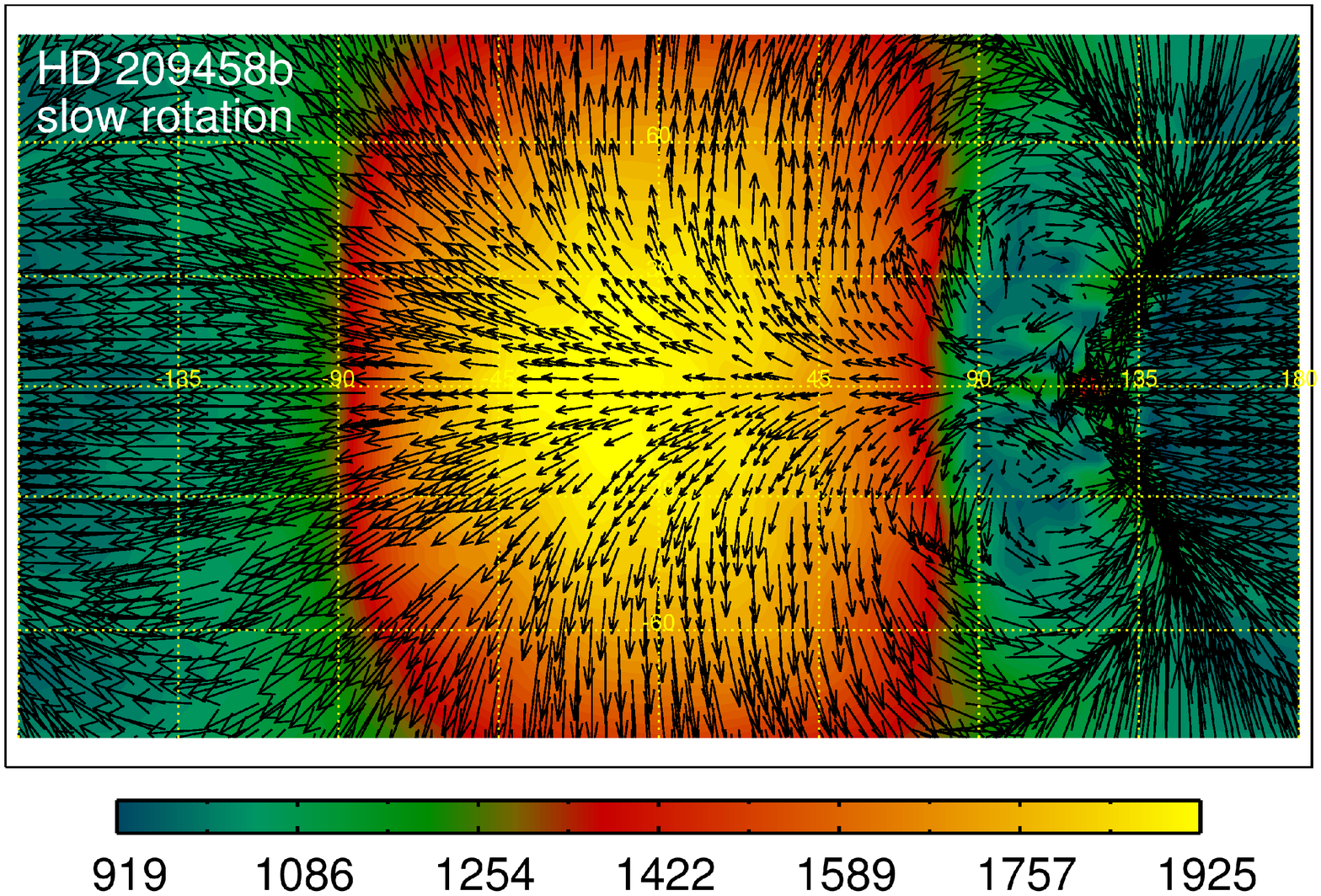} \\
\includegraphics[width=0.475\textwidth, trim= 0 0 0 60, clip=true]{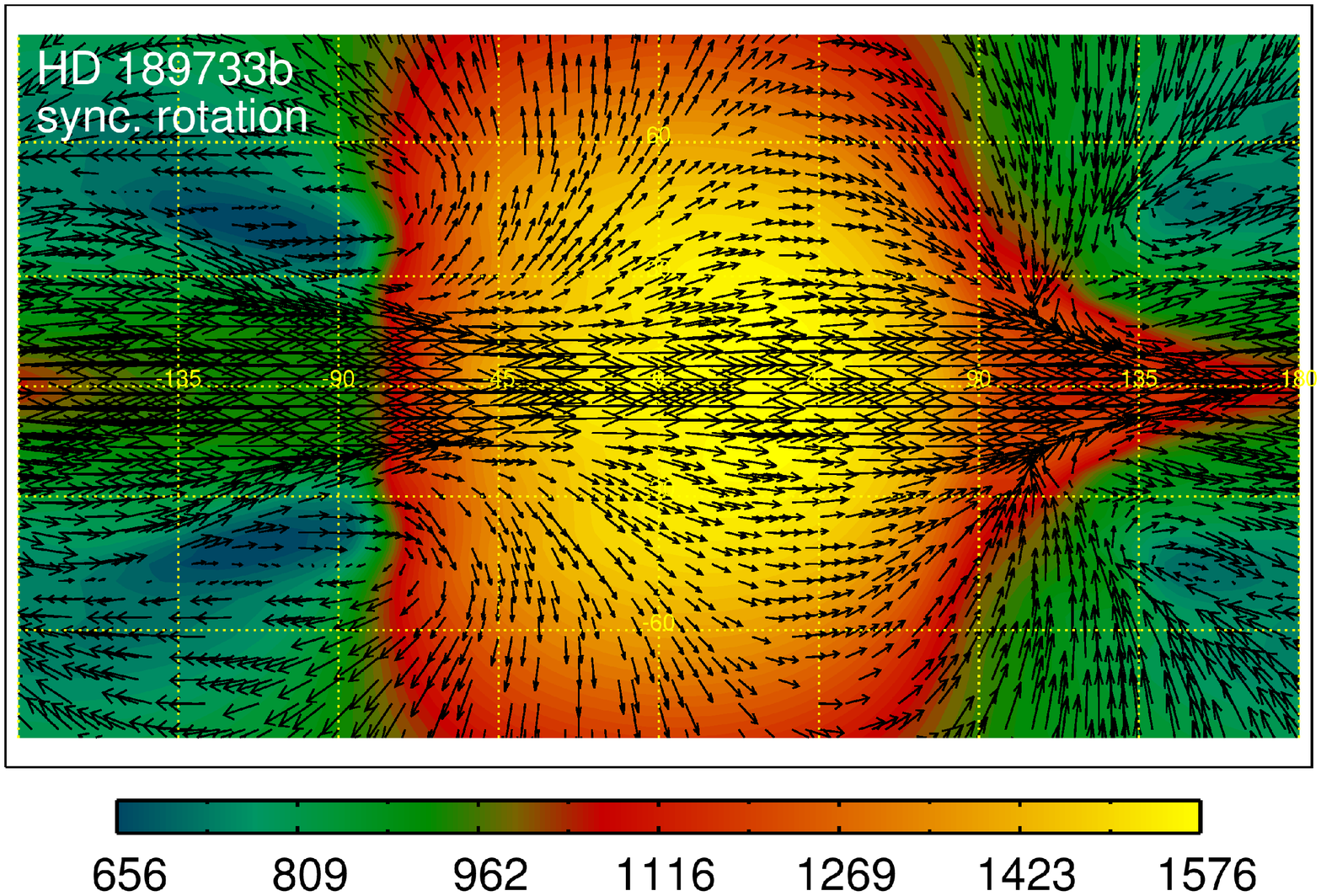}
\includegraphics[width=0.475\textwidth, trim= 0 0 0 60, clip=true]{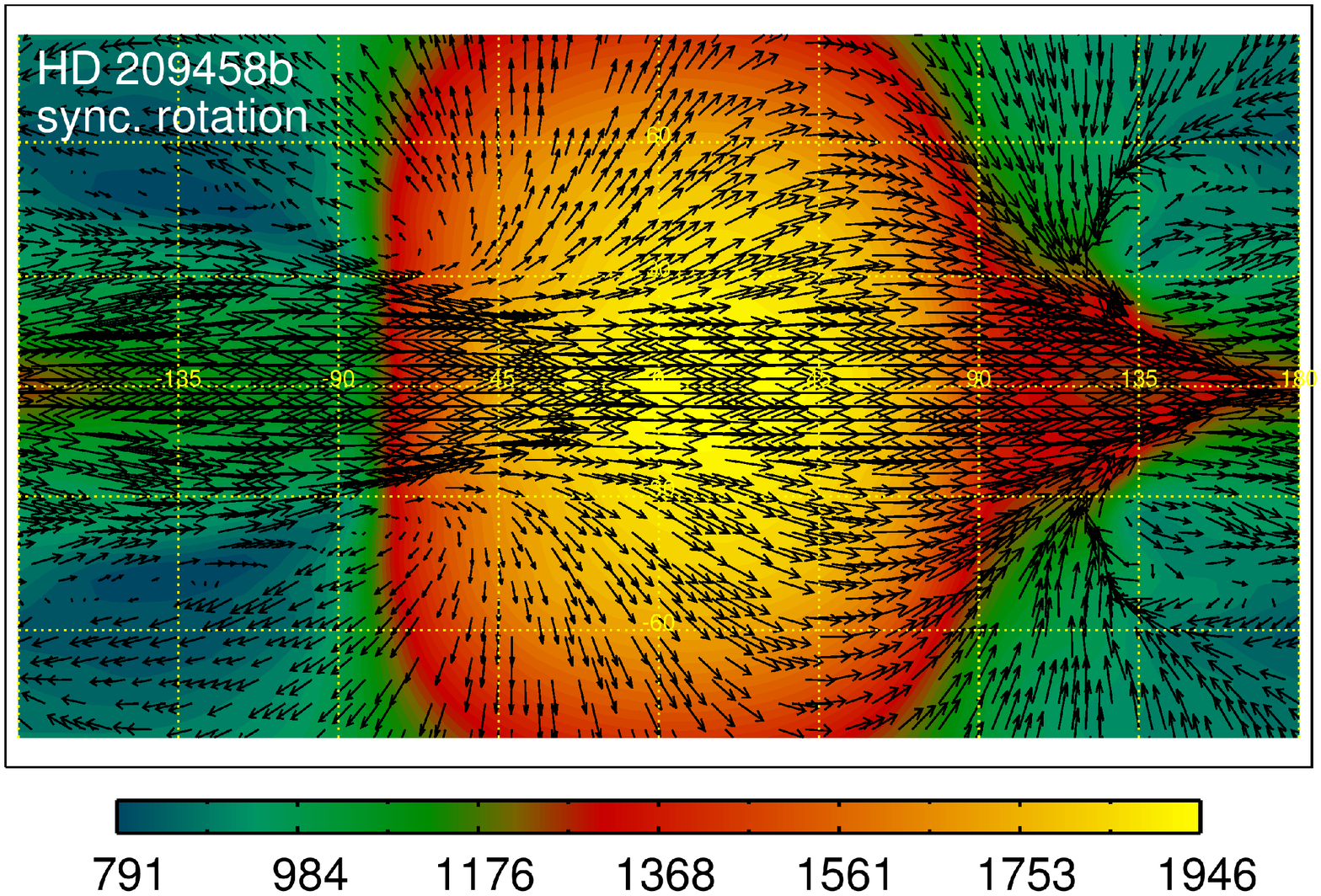} \\
\includegraphics[width=0.475\textwidth, trim= 0 0 0 60, clip=true]{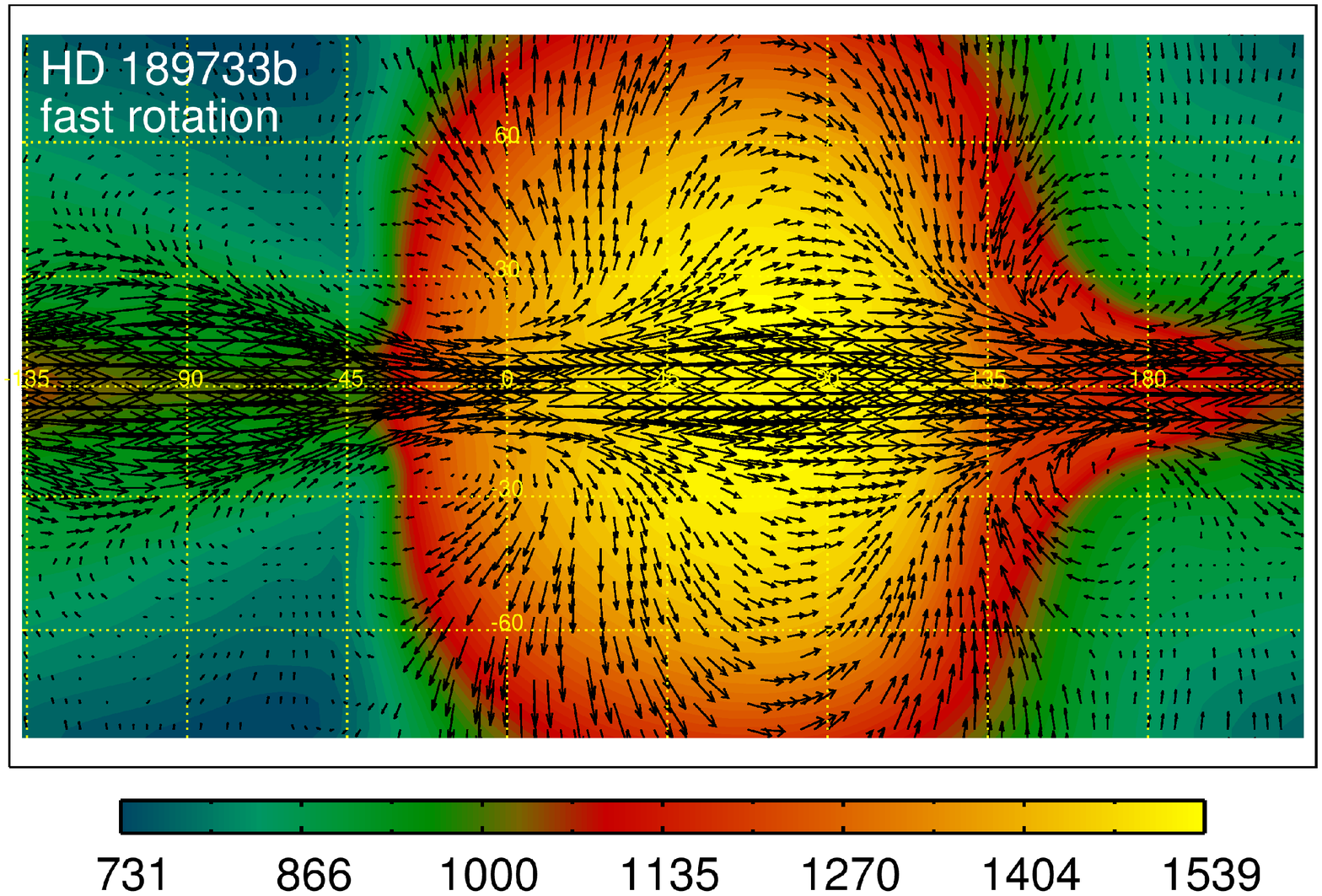}
\includegraphics[width=0.475\textwidth, trim= 0 0 0 60, clip=true]{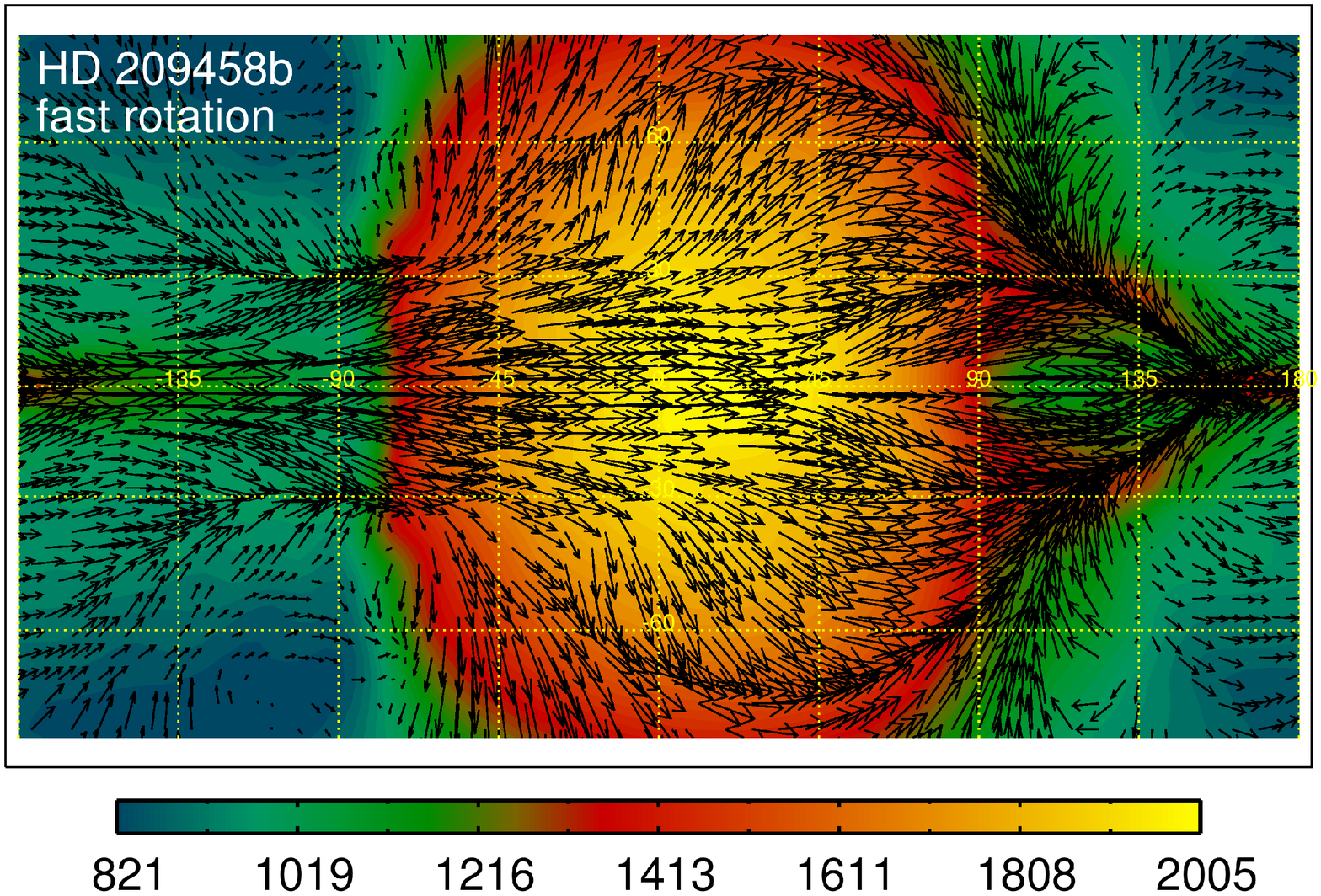}
\end{center}
\caption{Horizontal slices through the atmosphere, near the infrared photosphere, which is at 150 mbar for \HDo~and 53 mbar for \HDt.  Temperature is shown as color (in K) and winds as vectors, using a cylindrical projection centered on the substellar point.  From top to bottom, the maximum wind speeds for the \HDo~models are: 7, 6, and 5 \kms, and for the \HDt~models are: 6, 8, and 4 \kms, all measured in the frame rotating with the planet.} \label{fig:photmaps}
\end{figure}

Another way to compare the flow structures in each model is by viewing the temperature and wind patterns at the infrared photosphere, as shown in Figure~\ref{fig:photmaps}.  In all but the slowly rotating \HDt~model, the circulation pattern is characterized by an eastward jet along the equator and a temperature structure that has been advected away from a strict hot-day/cold-night pattern.  With increasing rotation rate the models develop a more significant high-latitude component of eastward flow on the day side of the planet.  In the zonal (east-west) averages shown in Figure~\ref{fig:uz} these components start to appear as the high-latitude eastward jets, but it is clear from these horizontal slices that they are limited to the day side, and there is sometimes a westward jet at these latitudes on the night side, in contrast to the global extent of the equatorial jet.

The slowly rotating \HDt~model, however, has a markedly different circulation pattern.  The flow is predominantly from day to night, over the western terminator.  The westward flow continues across the night side and is interrupted in a strongly convergent feature\footnote{Standard GCMs do not include treatments for accurately converting kinetic energy into thermal energy in shocks.  The general agreement between full Navier-Stokes models \citep{DobbsDixon2013} and models that use the primitive equations of meteorology (such as the one used here) may imply that shocks are not energetically or frictionally dominant in hot Jupiter atmospheres \citep[see also][]{Li2010}.  It is also not trivial to determine the extent to which hot Jupiter atmospheres should experience self-shocking, since there is no boundary against which the supersonic flow can shock \citep{Heng2012}.  Nevertheless, the importance of shocks in hot Jupiter atmospheres is an important topic that deserves further attention.} before reaching the day side.  Instead two vortices, stretching from the equator to each pole, circulate gas across the eastern terminator.  We have repeated the simulation for the slowly rotating \HDt~with slightly different initial conditions, to test against some kind of numerical error.  In all cases most of the atmosphere, above $\sim$0.1-1 bar, develops dominantly westward flow within the first few orbits of the planet.  The eddy momentum flux \citep[$\overline{u' v' \cos \phi}$, see Figure 12 of][and the surrounding discussion]{Kataria2013} of this model does show eastward flux away from the equator at near-photosphere pressures, in contrast to the eastward flux convergent on the equator in all other simulations.  

The mechanism responsible for the development of a super-rotating equatorial jet in hot Jupiter atmospheres was identified by \citet{Showman2011} and involves the differential propagation of planetary scale waves, a response of the atmosphere to the strongly asymmetric forcing pattern.  Their study was particular to the case of circular hot Jupiters in synchronous rotation, leading to the extreme and permanent day-night forcing pattern.  The jet-pumping mechanism, however, is robust against deviations from strict synchronicity.  This is shown in part by the non-synchronous models in \citet{Showman2009} and here, and even more so in the systematic study of GCMs for hot Jupiters on eccentric orbits by \citet{Kataria2013}, in which the authors presented a range of models for eccentric planets in pseudo-synchronous rotation and found ubiquitous super-rotating equatorial jets, with the changes in jet structure as expected for varying rotation rate.  It is beyond the scope of this paper to provide a detailed analysis of the regime shift observed in the slowly rotating model of \HDt, but here we comment upon some of the parameters that are likely to be important, namely: the rotation rate, other relevant atmospheric timescales, and the relative motion of the substellar point.

To further investigate the atmospheric circulation regime of slowly rotating planets we ran additional simulations (not shown here, but to be presented in detail in future work): a model of \HDo~at the same slow rotation rate as \HDt~($\Omega_{\mathrm{rot}} = 1.05 \times 10^{-5}$ s$^{-1}$), and two more of \HDt~with different slow rotation rates.  In all cases the simulations failed to produce super-rotating equatorial jets, although the circulation patterns in each model were different, including the acceleration of winds and time evolution of the flow.  Our models of \HDt~and \HDo~at $\Omega_{\mathrm{rot}} = 1.05 \times 10^{-5}$ s$^{-1}$ both produce a photospheric temperature structure with the hottest region to the west of the substellar point; however, the flow patterns differ between the two models.  The model of \HDt~has coherent westward flow at this pressure level (see Figure~\ref{fig:photmaps}), while the model of \HDo~has a component of eastward flow from the hottest region, across the substellar point, and continuing just over the terminator until it encounters the westward flow on the night side.  This indicates that the rotation rate is important in the producing the circulation regime shift, but is not the sole parameter that controls the circulation and temperature patterns, as is to be expected.\footnote{This is immediately obvious, for example, if we compare our models of hot Jupiter atmospheres to slowly rotating models of hypothetical terrestrial (or water-world) exoplanets  \citep[e.g.,][]{Walker2006,Merlis2010,Edson2011,Yang2013} or the atmosphere of Venus ($P_{\mathrm{rot}}=243$ days).  The presence of a surface is the dominant difference in such a comparison and strongly influences the nature of the atmospheric circulation.}

A quick comparison of the timescales relevant to the heating and circulation of the atmosphere provides no glaring difference between models, no obvious reason for the regime shift.  In Table~\ref{tab:taus} we calculate the timescales of: radiative heating \citep[$\tau_{\mathrm{rad}}=c_p P/4\sigma T^3 g$, e.g.][]{SCM2010}, gravity wave propagation \citep[$R_p/\sqrt{gH}$,][]{PerezBecker2013}, rotation ($\omrot^{-1}$), and the movement of the substellar point ($\mid \omorb-\omrot \mid^{-1}$), using the parameters given in Table~\ref{tab:params}.\footnote{Another timescale relevant to atmospheric circulation is one related to sources of drag \citep{PerezBecker2013}, but as we have not included any explicit drag in our model, we do not include a drag timescale in Table~\ref{tab:taus}.}  While the rotational timescale for the slowly rotating \HDt~is clearly much longer than that planet's radiative and wave timescales, and by a larger factor than the difference in timescales for \HDo, this does not immediately lead to a simple explanation as to why we should see a circulation regime shift for this model.

\begin{deluxetable}{lcccc}
\tablewidth{0pt}
\tablecaption{Atmospheric timescales (in seconds)}
\tablehead{
\colhead{Model}  &  \colhead{Radiative heating\tablenotemark{a}} & \colhead{Wave propagation\tablenotemark{b}}
  & \colhead{Rotation}  & \colhead{Substellar motion}
}
\startdata
\HDo~slow  &    $2\times10^{4}$ & $4\times10^{4}$ &  $6\times10^{4}$ & $6\times10^{4}$ \\
\HDo~locked  & $2\times10^{4}$ & $4\times10^{4}$ &  $3\times10^{4}$ & $\infty$ \\
\HDo~fast  &     $2\times10^{4}$ &  $4\times10^{4}$ & $2\times10^{4}$ & $3\times10^{4}$ \\
\HDt~slow  &     $1\times10^{4}$ & $4\times10^{4}$ &  $1\times10^{5}$ & $1\times10^{5}$ \\
\HDt~locked  &  $1\times10^{4}$ & $4\times10^{4}$ & $5\times10^{4}$  & $\infty$ \\
\HDt~fast  &      $1\times10^{4}$ &  $4\times10^{4}$ & $2\times10^{4}$ & $8\times10^{4}$ \\
\enddata
\tablenotetext{a}{We calculate this value at the infrared photosphere of each model, using $T=1200, 1500$ K for \HDo, \HDt.}
\tablenotetext{b}{The scale height of the atmosphere is $H=kT/mg$ and we assume that molecular hydrogen is the primary constinuent of the atmosphere.  We again use $T=1200, 1500$ K for \HDo, \HDt.}
\tablecomments{See the text for definitions and a discussion of these timescales.}
\label{tab:taus}
\end{deluxetable}
\clearpage

Not reflected in these timescales is the directionality of motion.  Rotation defines the eastward (positive) direction, which is also the direction that equatorial Kelvin waves propagate \citep{Showman2011}.  In the slowly rotating models the substellar point moves to the east, while it moves west in the quickly rotating models.  One might suppose that a relevant parameter would be the thermal lag of the atmosphere in response to the moving substellar point, which we could characterize as an offset in distance: $d_{\mathrm{lag}}=-R_p (d\tss/dt) \tau_{\mathrm{rad}}$.  However, it is easy to show that the ratio of this parameter for the slowly rotating models of \HDo~and \HDt~is: $d_{\mathrm{lag,HD1}}/d_{\mathrm{lag,HD2}}=(T_{\mathrm{HD2}}/T_{\mathrm{HD1}})^3 (\omega_{\mathrm{orb, HD1}}/\omega_{\mathrm{orb, HD2}}) \approx 3$, if defined in units of each planet's radius.  This means that the atmosphere of \HDo~experiences a much larger thermal lag shift in response to the eastward motion of the substellar point, making this parameter an unlikely culprit for the change in circulation pattern on \HDt.

Clearly a simple explanation of this regime shift can not be made through a comparison of rotation rates, atmospheric timescales, or the motion of the substellar point.  While a precise identification of the cause for the regime shift must be left for a more detailed analysis in future work, it does seem to be a robust feature of the very slow rotation rate for \HDt~and has directly observable consequences, as discussed in the following section.

\section{The effect of rotation on observable properties} \label{sec:obs}

Currently the only observational method for directly constraining circulation patterns on hot Jupiters is by measuring the spatial variation of their thermal emission, either through orbital phase curves or eclipse maps of their daysides (see above).  While Doppler measurements of hot Jupiter orbital motion is currently achievable, the additional Doppler signal from winds and rotation is just shy of being measured to statistical significance \citep[e.g.,][]{Snellen2010,Brogi2012,Rodeler2012,Birkby2013,deKok2013a}.  Nevertheless, it is our expectation that near-future improvements in instrumentation or methodology may make these Doppler shifts observable \citep{deKok2013b}, and this method should become more widely applicable in the era of 30-m class telescopes.  Therefore, here we compare the signatures of planetary rotation in thermal emission and Doppler-shifted transit spectra, in order to understand whether these observational methods could be used together to constrain the rotation rate of a planet.

\subsection{Thermal emission, observed as orbital phase curves}

The temperature structure at the infrared photosphere is a good indication of what sort of properties may be observed in thermal emission from the planet, but one product of our GCM is a spatial map of the actual flux emitted from the planet, which we can integrate over the observed hemisphere as a function of time to produce predicted orbital phase curves, as described in more detail above.  Based on the temperature structures shown for each model in Figure~\ref{fig:photmaps}, we expect a fair amount of variation in flux between the colder night sides and hotter day sides, and more so for the planet \HDt, which has a larger day-night temperature difference.  We also expect that in all models but the slowly rotating \HDt, we should see the flux peak before the substellar point is directed at the observer, since the temperature maximum is to the east of the substellar point; in the slowly rotating model of \HDt~the hottest region has been advected to the west and we expect the phase curve should peak later.  However, the non-synchronous rotation of the planet complicates our expectations somewhat, since the observed thermal emission is a function of both the advected temperature structure and the rotation rate \citep[for example, the semi-analytic models of][recognize that the rotation rate of the planet and wind speeds are observationally degenerate]{Cowan2011a}.

\begin{figure}[ht!]
\begin{center}
\includegraphics[width=0.16\textwidth]{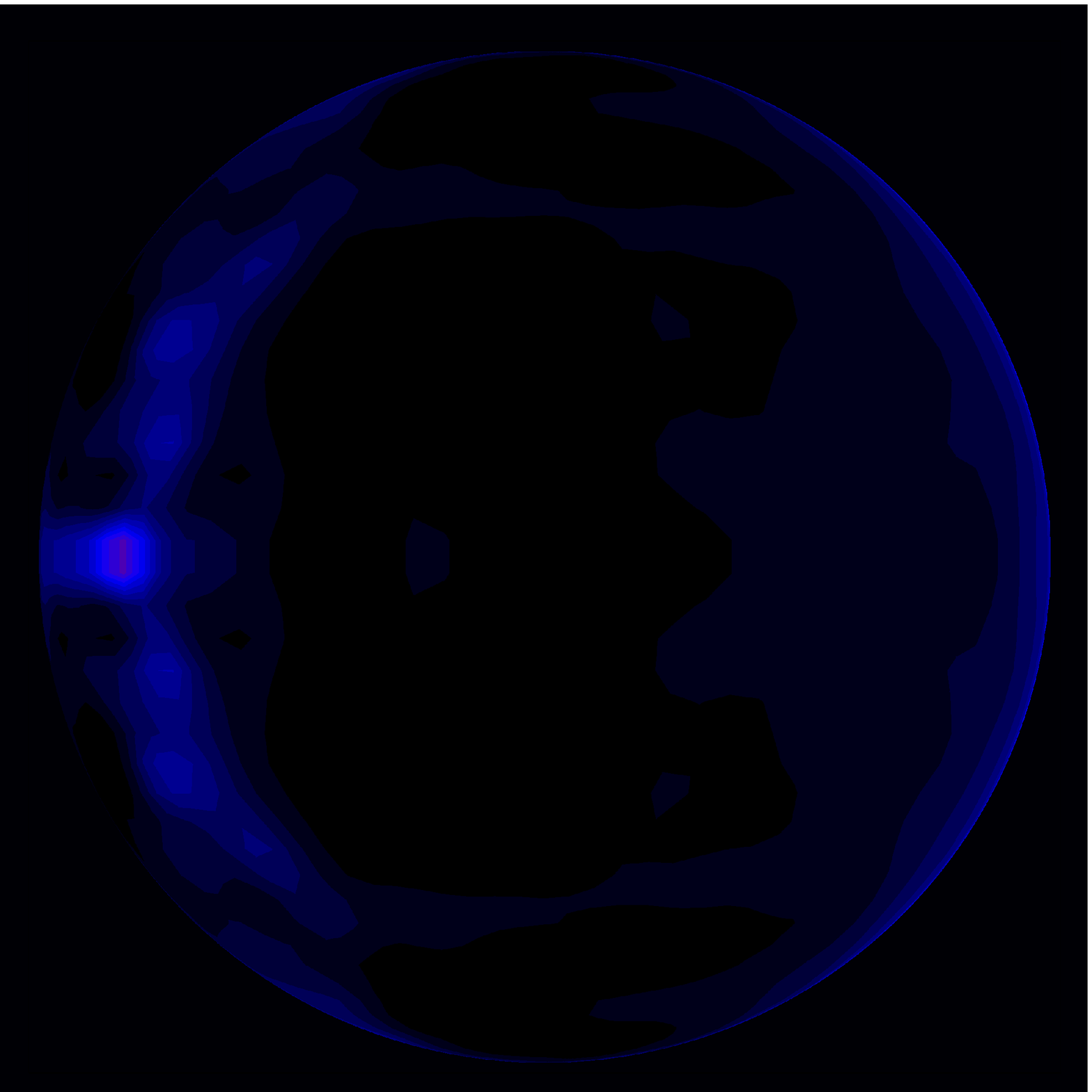}
\includegraphics[width=0.16\textwidth]{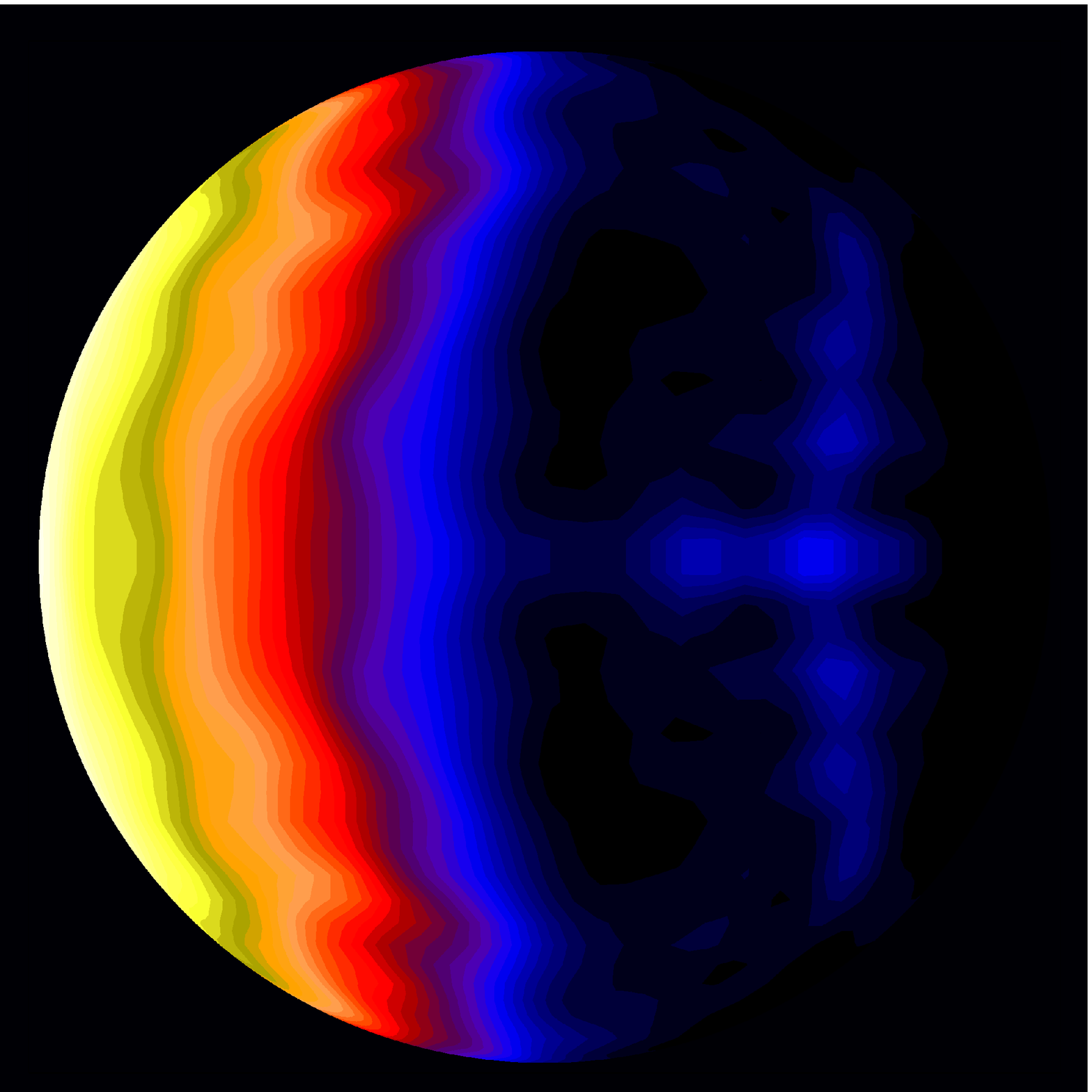}
\includegraphics[width=0.16\textwidth]{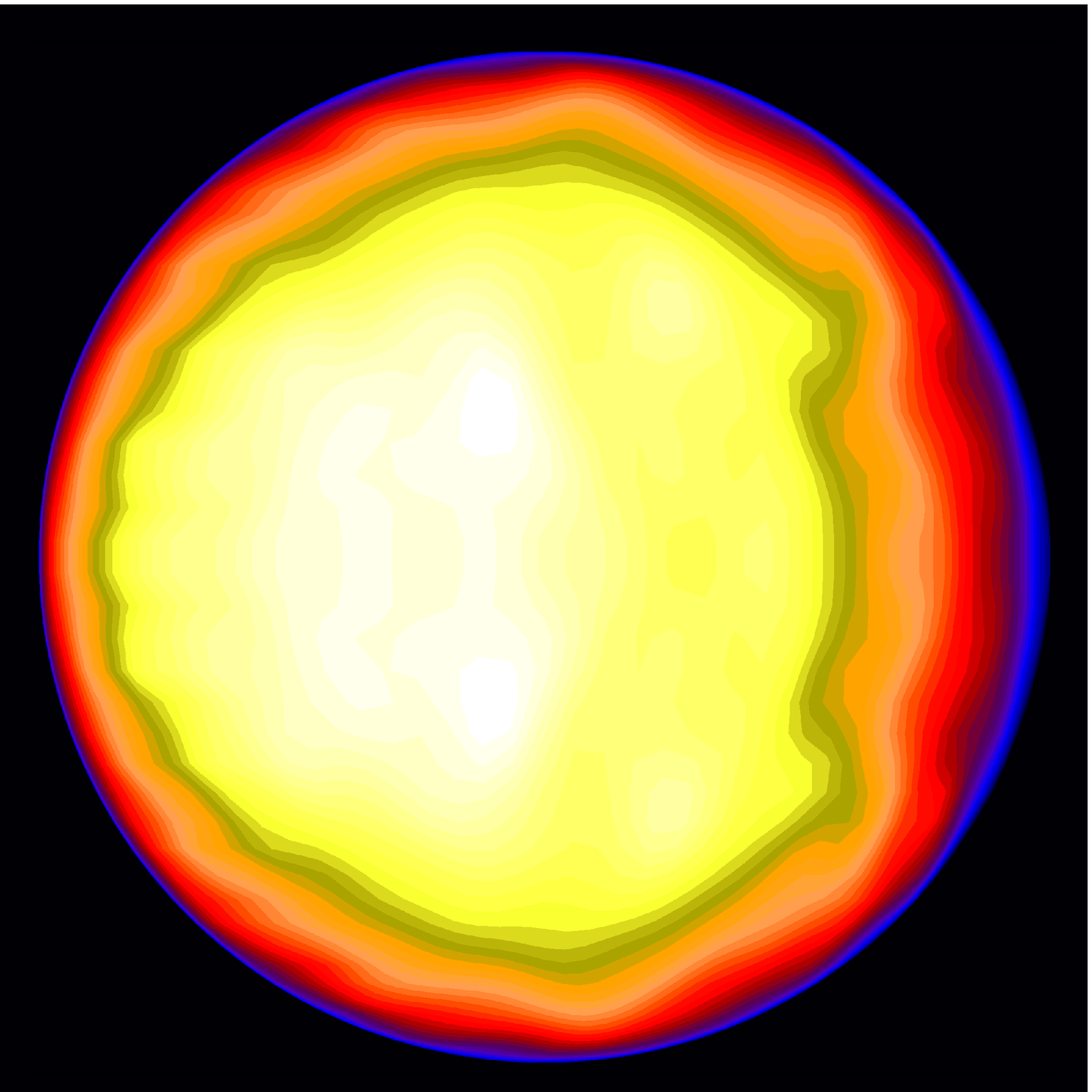}
\includegraphics[width=0.16\textwidth]{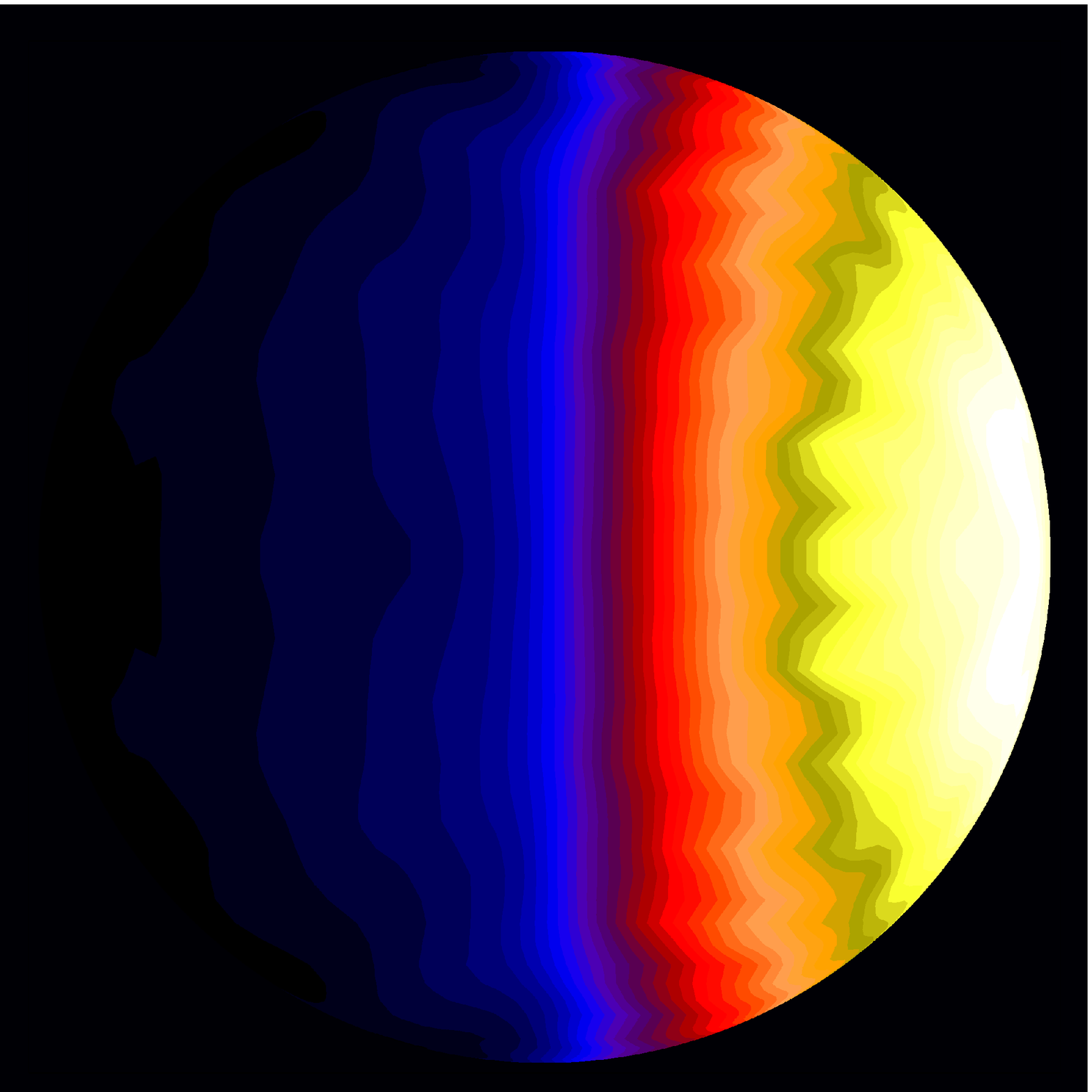}
\includegraphics[width=0.16\textwidth]{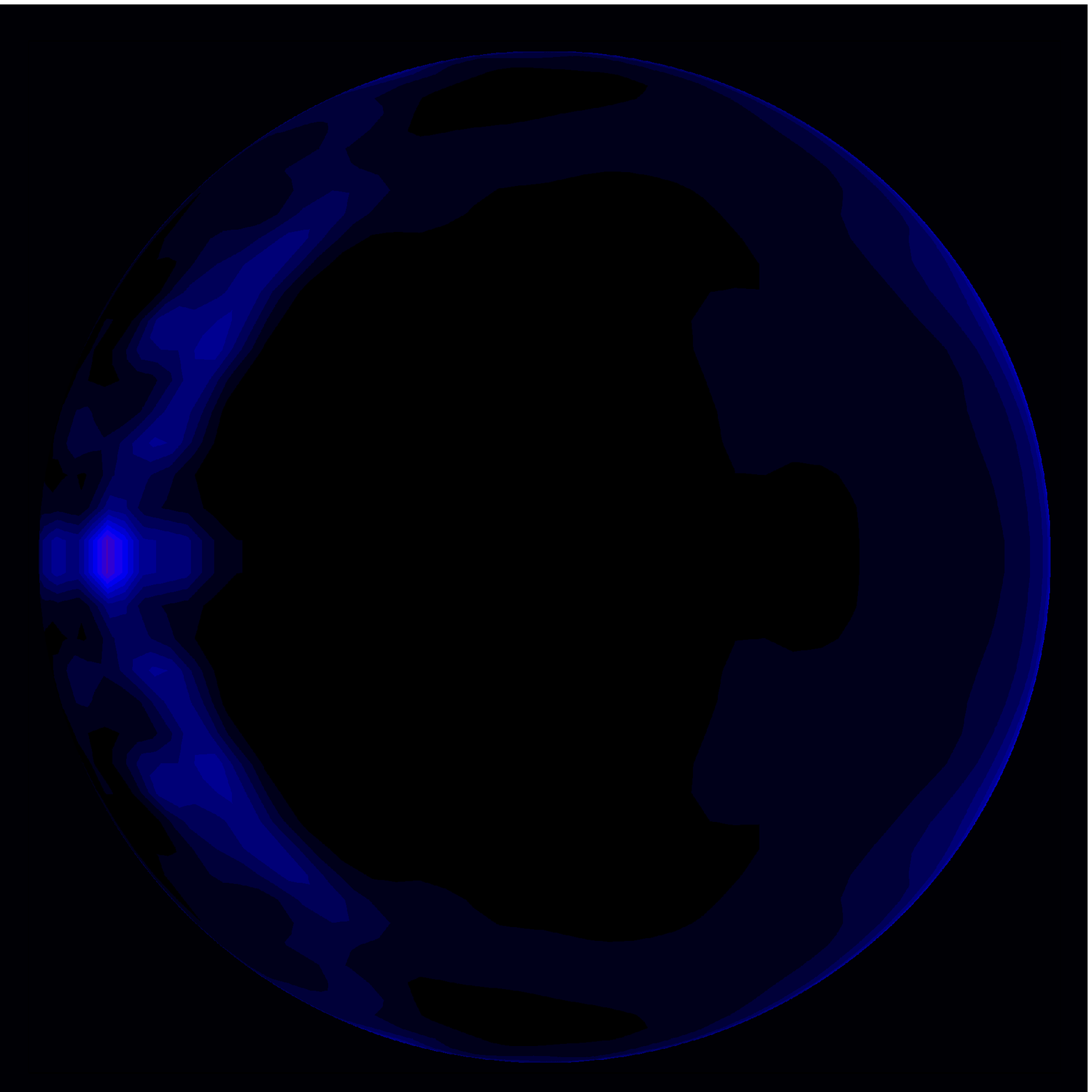} \\
\includegraphics[width=0.16\textwidth]{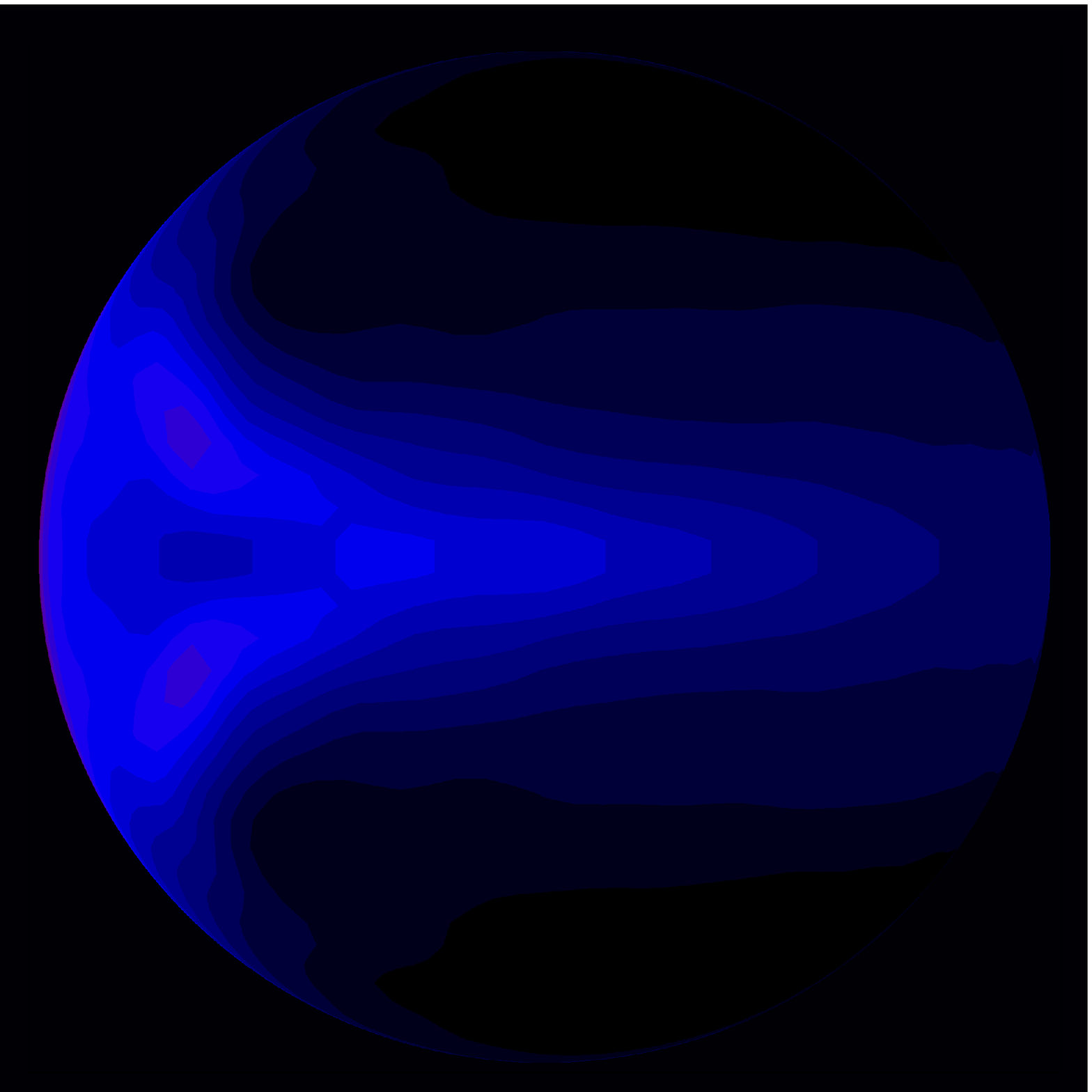}
\includegraphics[width=0.16\textwidth]{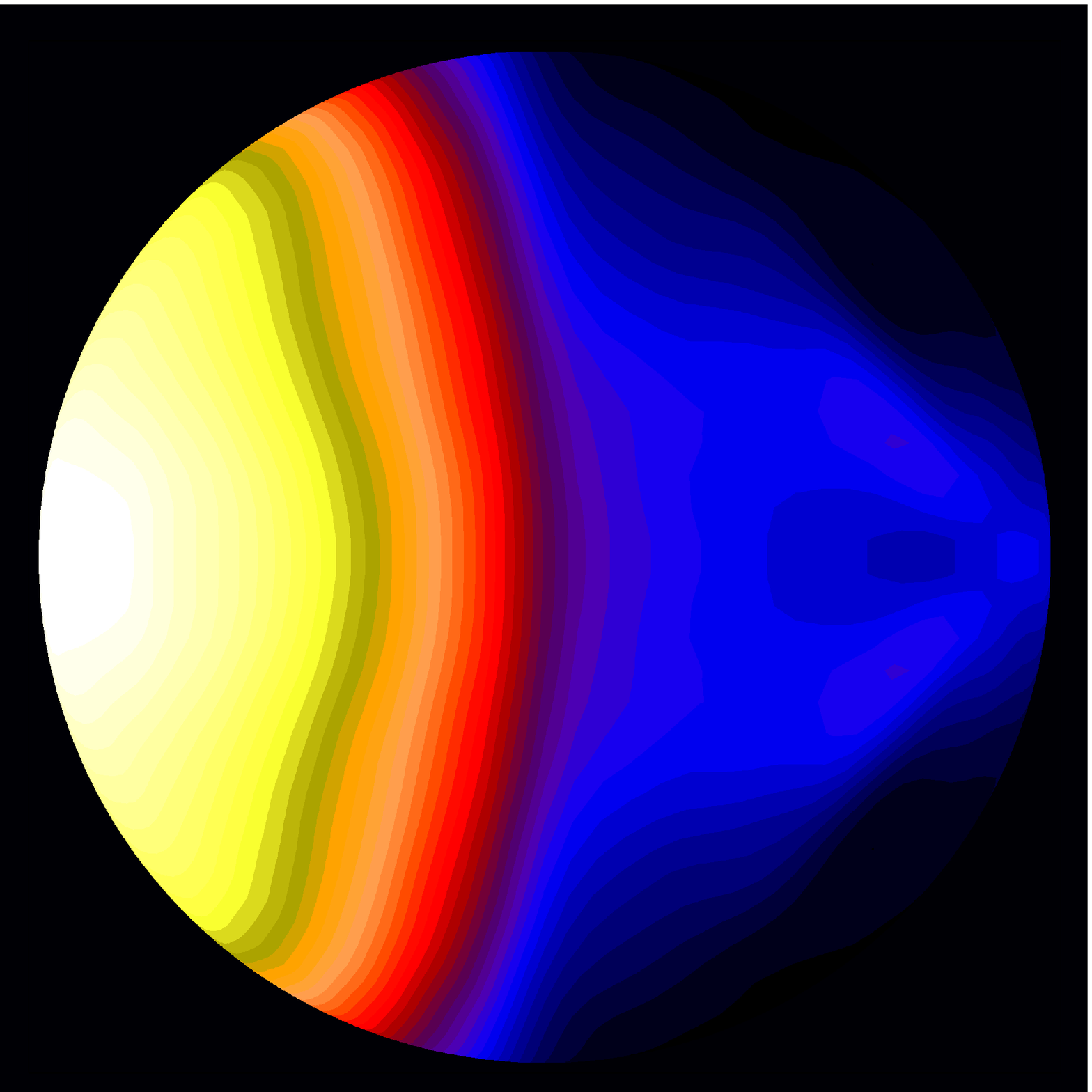}
\includegraphics[width=0.16\textwidth]{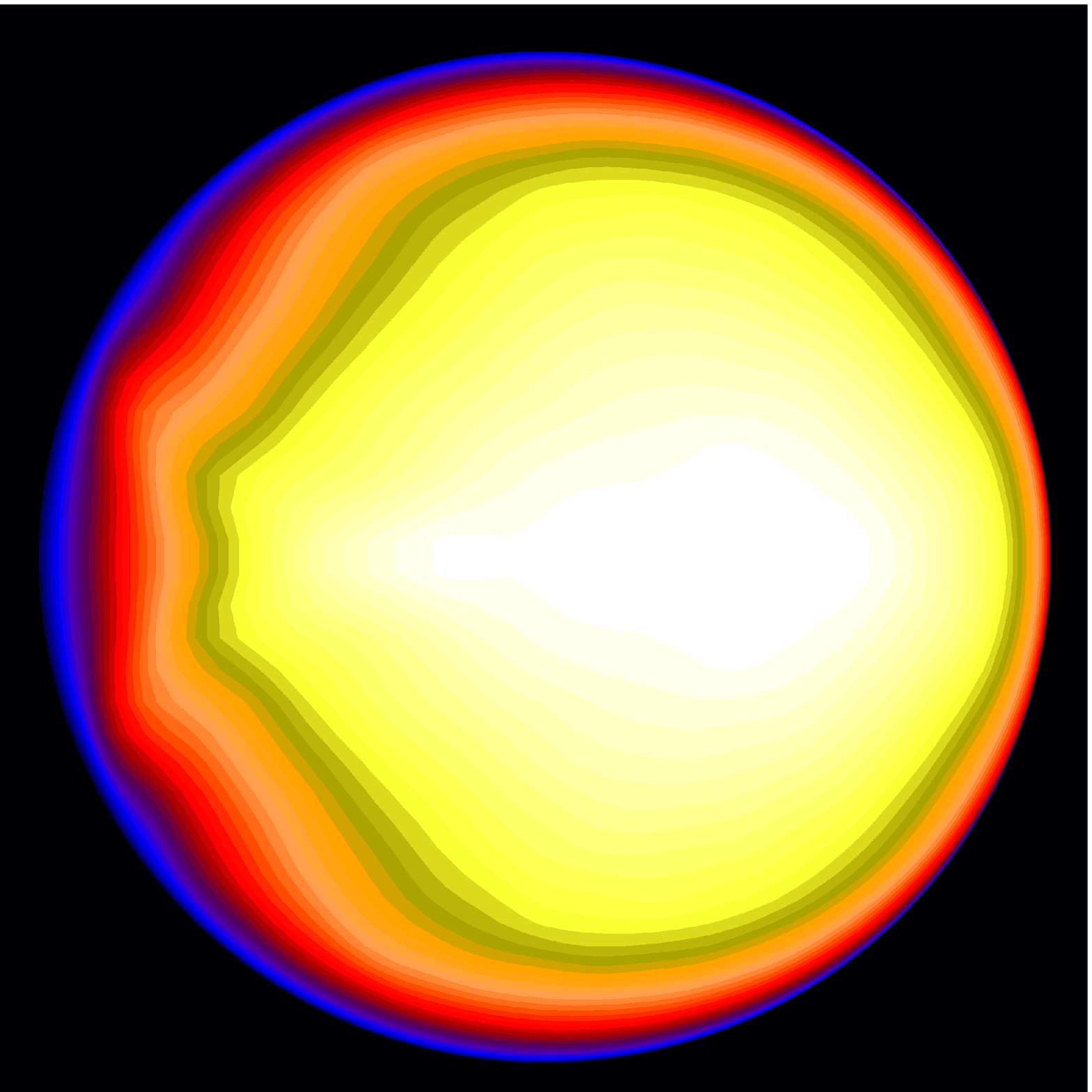}
\includegraphics[width=0.16\textwidth]{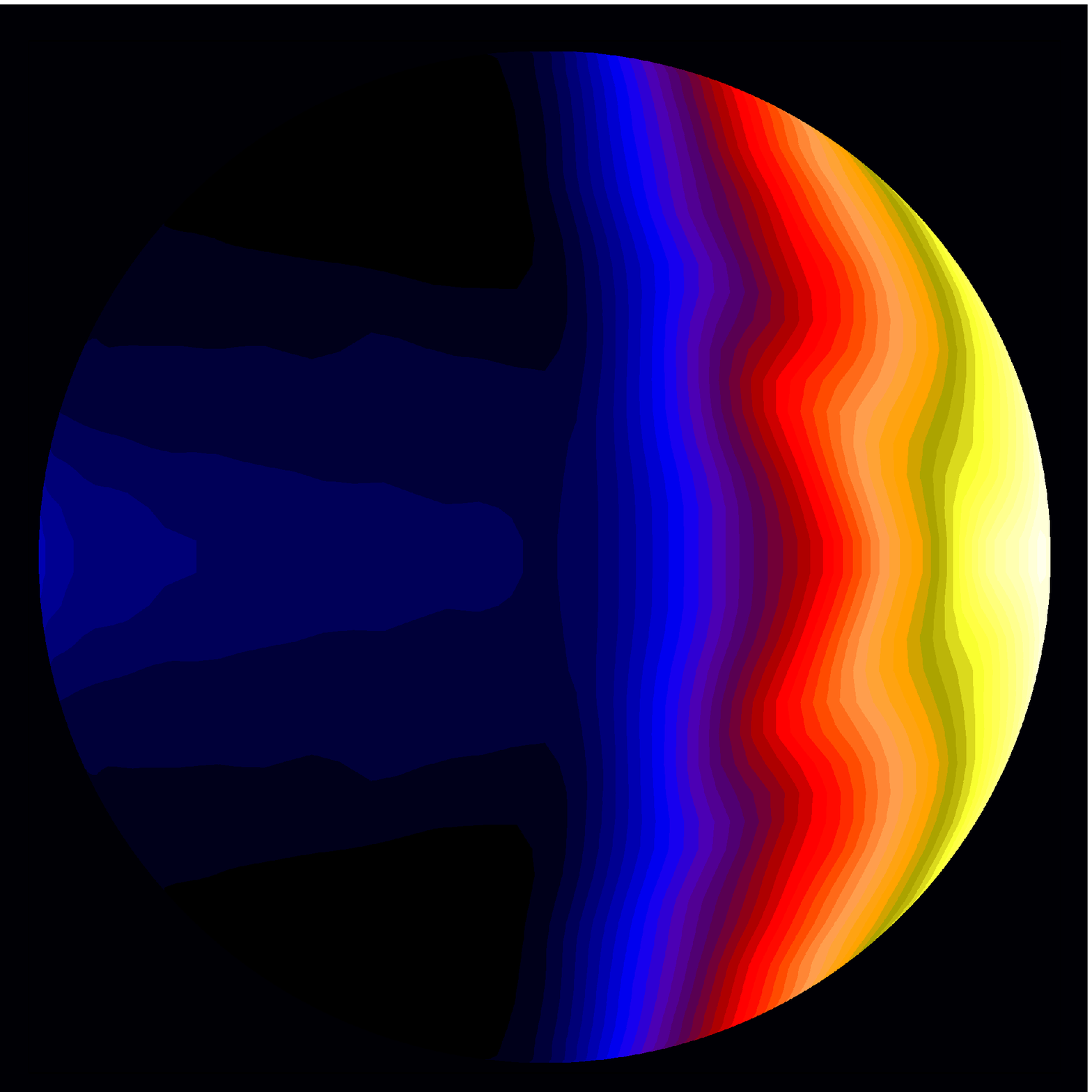}
\includegraphics[width=0.16\textwidth]{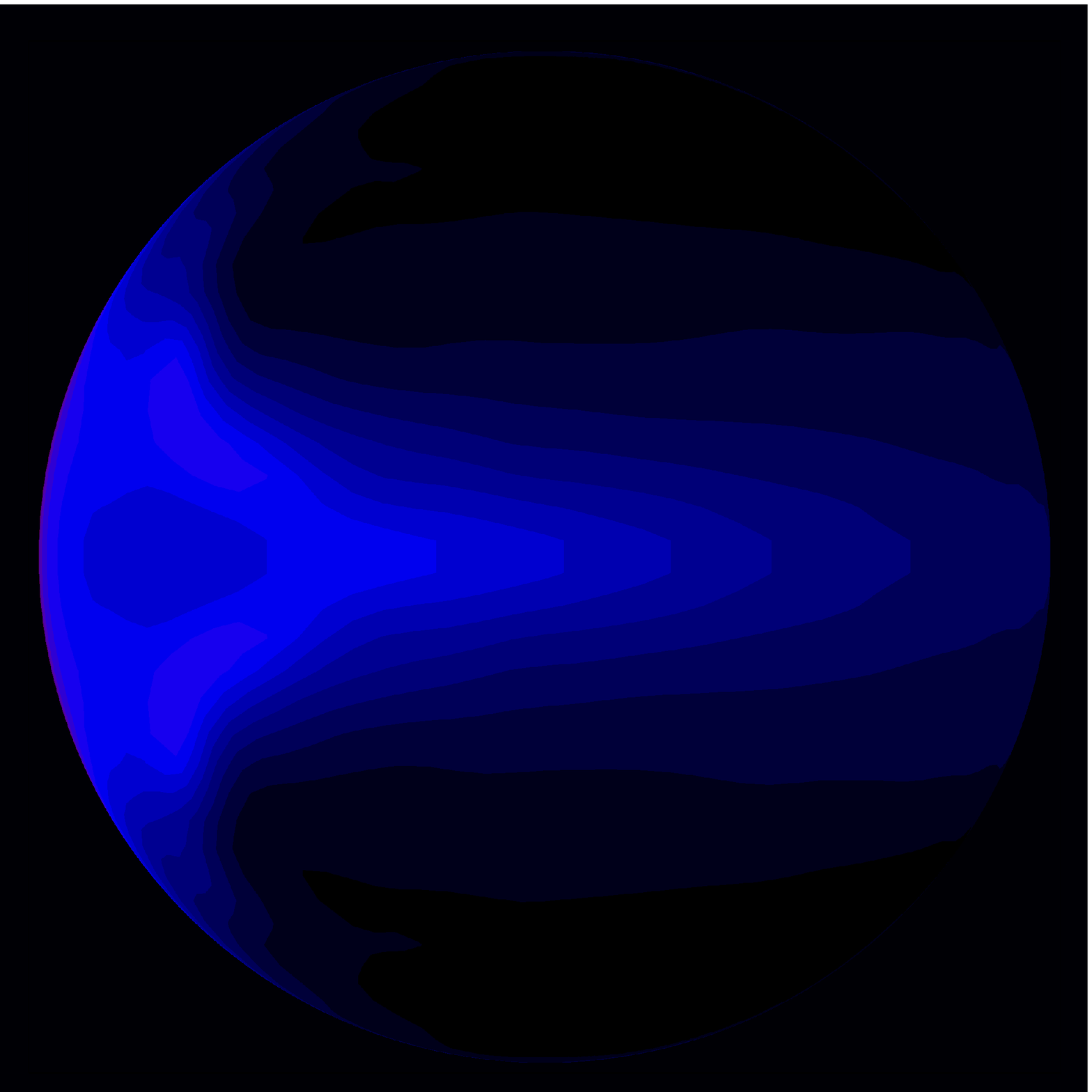} \\
\includegraphics[width=0.16\textwidth]{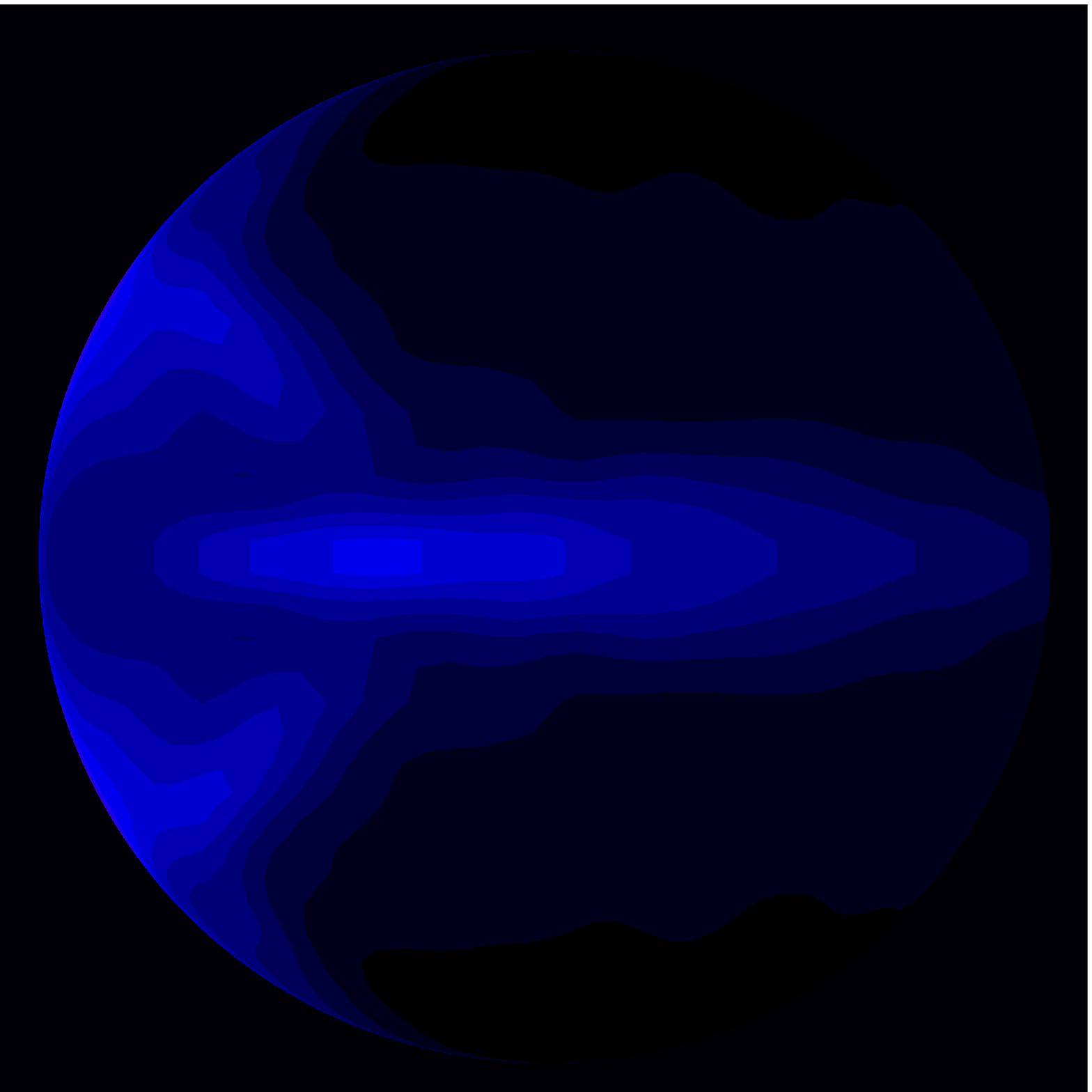}
\includegraphics[width=0.16\textwidth]{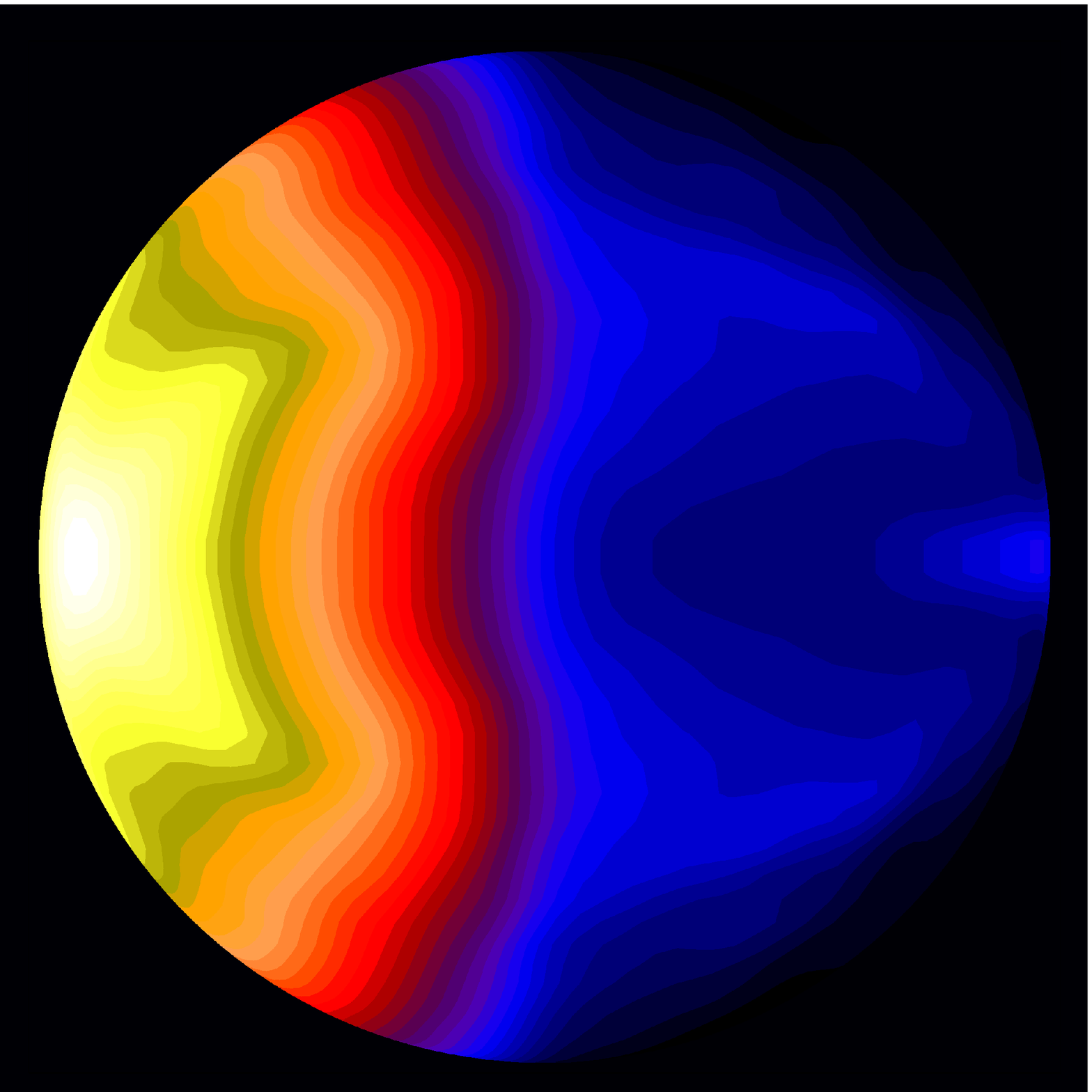}
\includegraphics[width=0.16\textwidth]{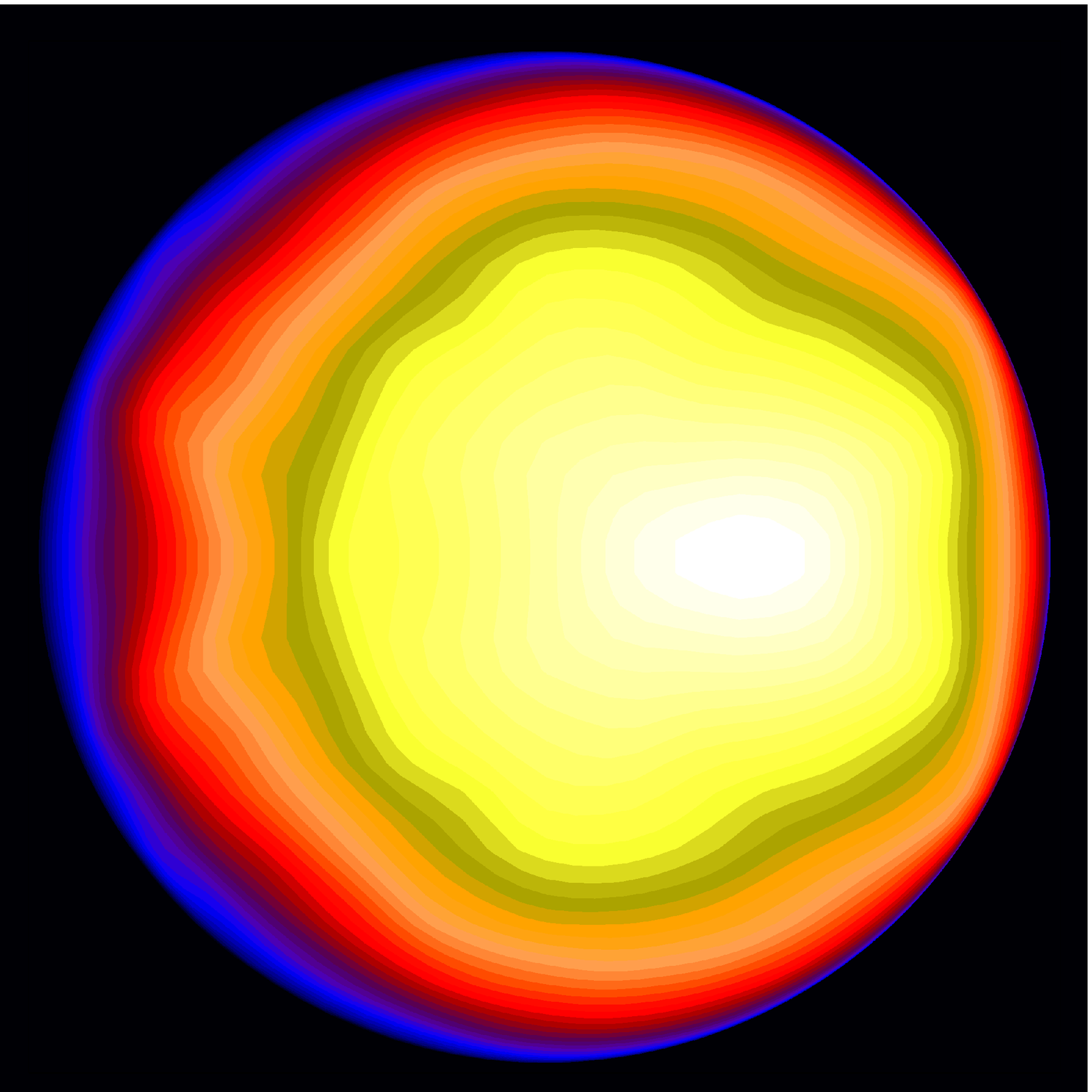}
\includegraphics[width=0.16\textwidth]{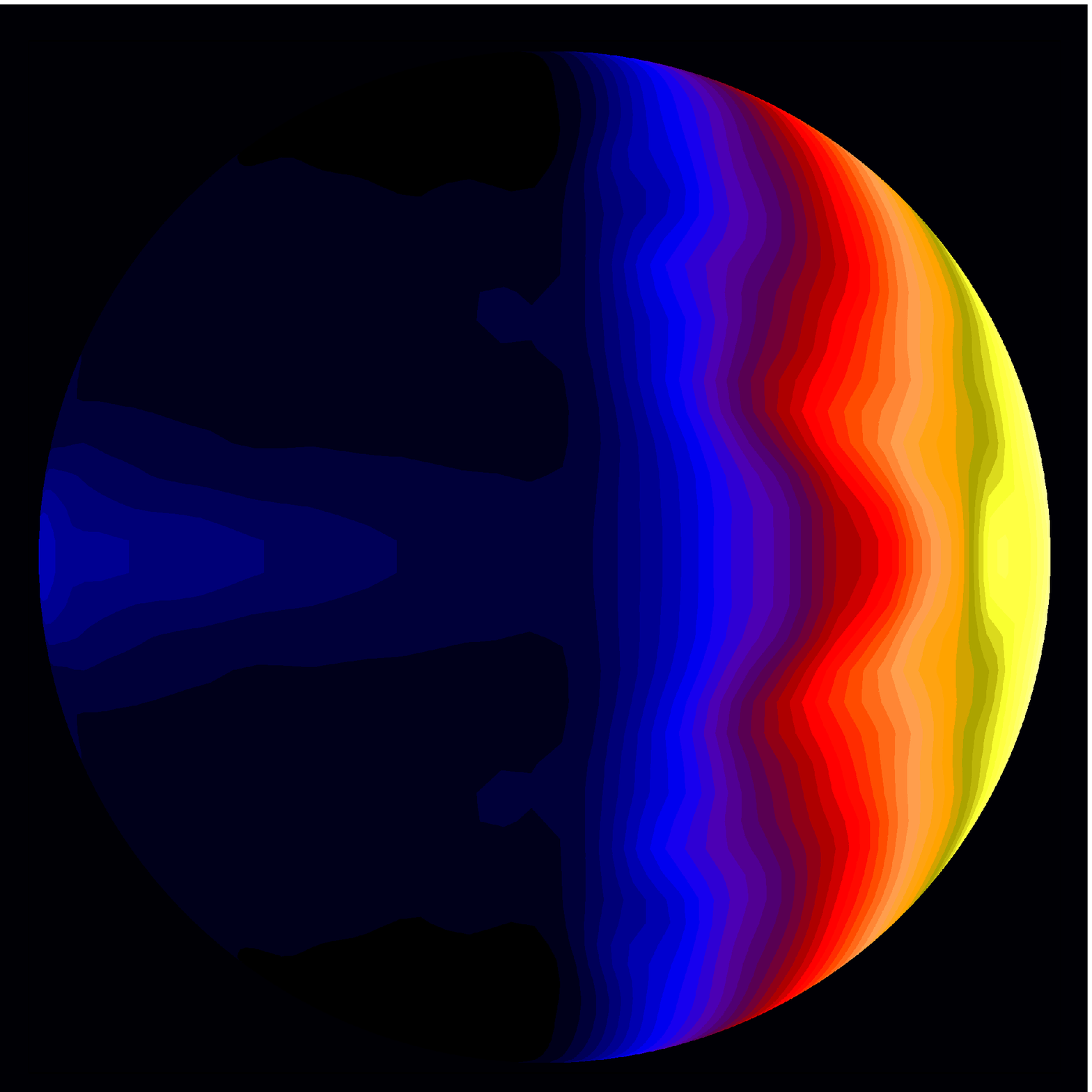}
\includegraphics[width=0.16\textwidth]{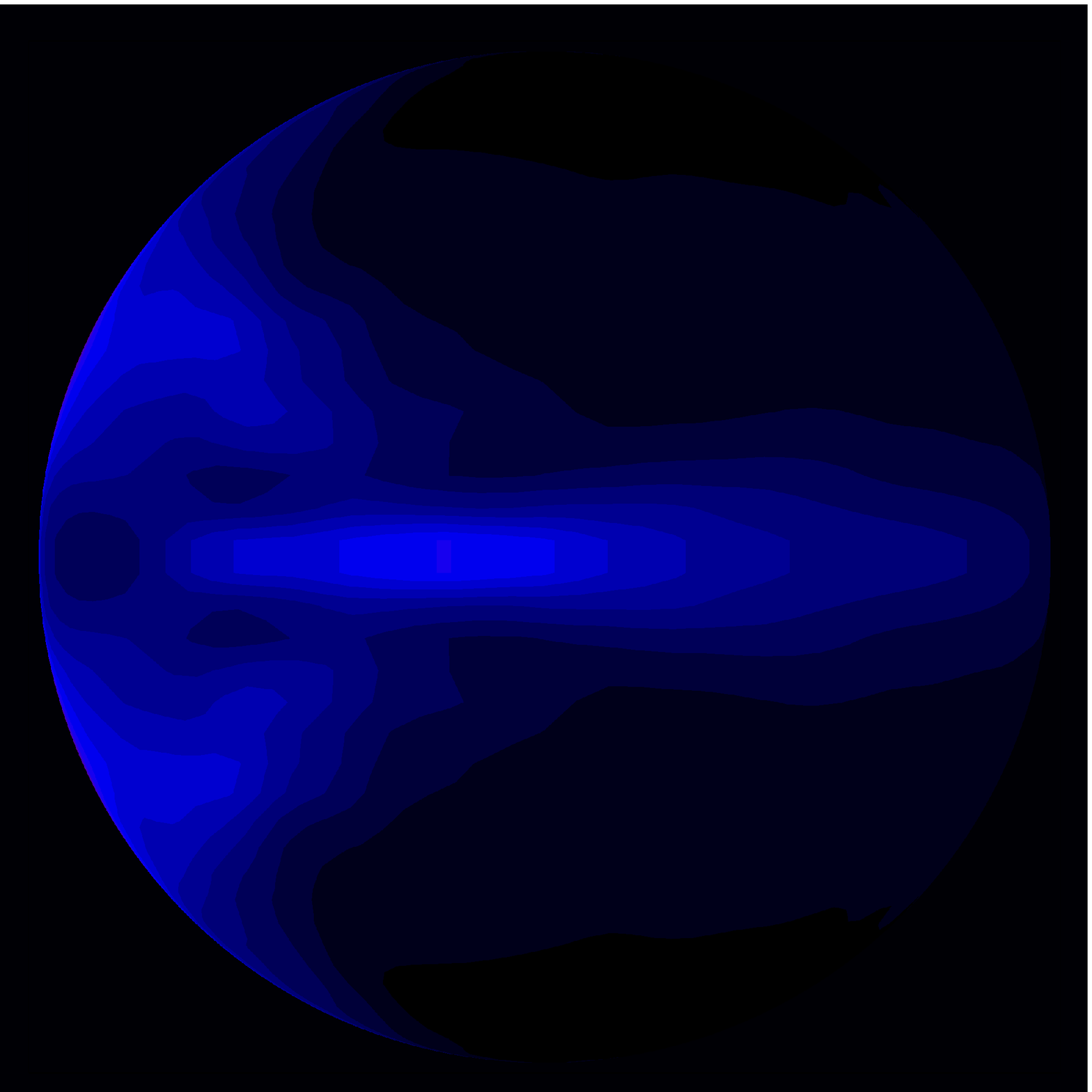}
\includegraphics[width=0.9\textwidth]{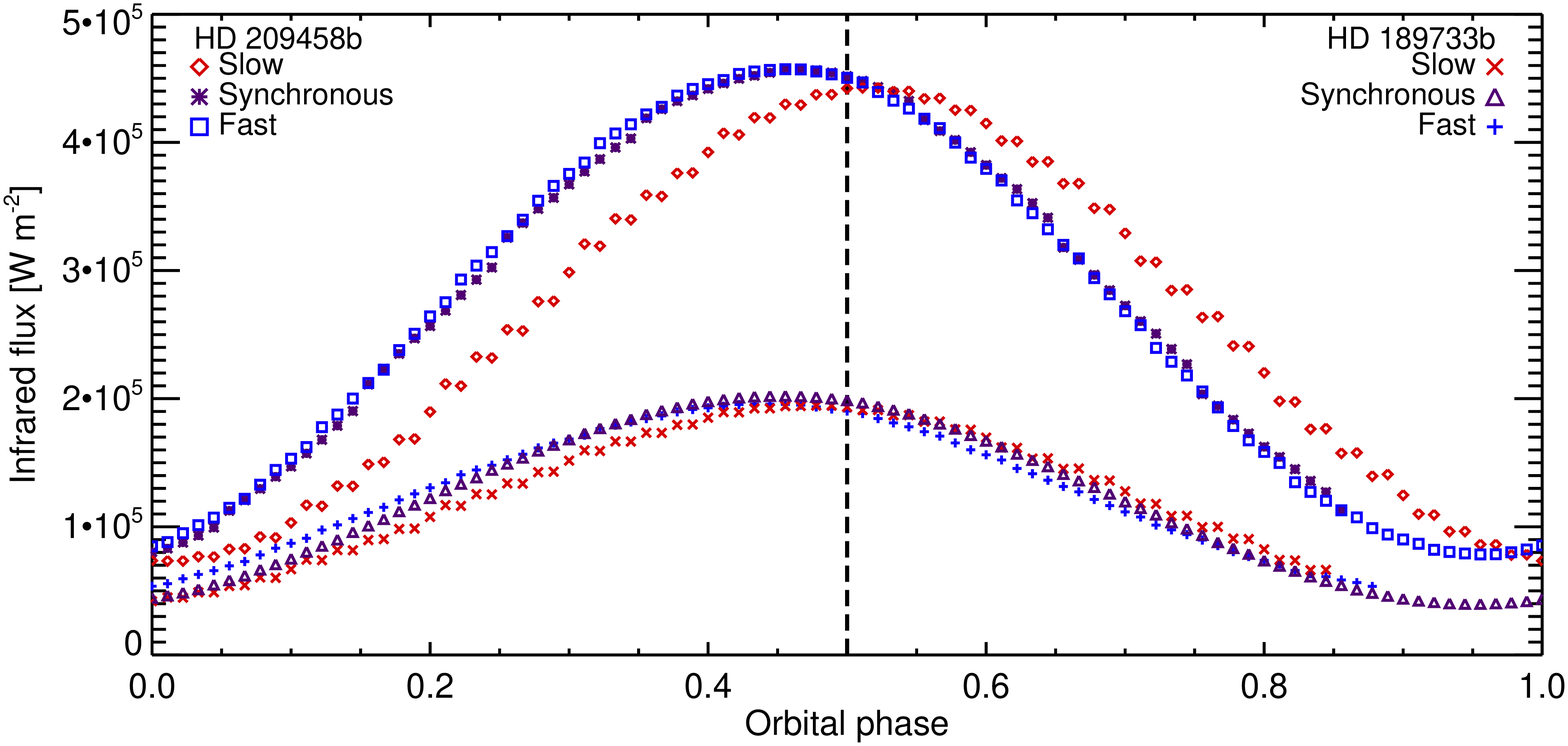}
\end{center}
\caption{The infrared emission that would be observed throughout a planet's orbit.  The color images show maps of \HDt's emitted flux at regular intervals during a single orbit; \emph{from left to right} the snapshots run from inferior conjunction to superior conjunction and back around to the starting point.  The \emph{top, middle, and bottom} rows show the models with slow, synchronous, and fast rotation, respectively.  The bottom plot shows the flux from the hemisphere oriented toward the observer, for each model.  The dashed line at 0.5 phase marks where the secondary eclipse would occur, which is also where the phase curve would peak for a planet with a hot spot aligned with the substellar point.}\label{fig:pc}
\end{figure}

The orbital phase curves from each of our simulations is shown in Figure~\ref{fig:pc}, where we also show flux images of the \HDt~models for several snapshots throughout a single orbit.  These phase curves demonstrate the observational degeneracy between the rotation rate and the advected shift of the hottest region of the atmosphere; the models of \HDo, in particular, have very similar temperature structures, but the most quickly rotating model peaks slightly sooner than the synchronous one, which itself peaks sooner than the slowly rotating model.  This is the same trend found for the non-synchronous \HDo~models in \citet{Showman2009}, with the differences between models comparable to the differences between models that varied other poorly constrained parameters (such as the atmospheric metallicity).

Although we confirm the finding from \citet{Showman2009}, that the rotation rate of \HDo~cannot be unambiguously determined from its orbital phase curve, we find a strong contrast between the fast and synchronously rotating models of \HDt~and its slowly rotating model.  Since the synchronous rotation rate of \HDt~is slower than \HDo, we predict that the regime transition we find for models of \HDt~rotating more slowly than synchronous should result in an orbital phase curve that peaks after, rather than before, secondary eclipse (when the substellar point directly faces the observer).  This delayed peak is significantly distinct from the earlier peak seen in all of our other models.  

To our knowledge, the only other published instance of a predicted phase curve that occurs later than secondary eclipse is from the highly variable atmospheric models in \citet{Rauscher2008}, from the simulations of \citet{Cho2003,Cho2008}.  Those models predicted a completely different temperature structure, characterized by large, cold vortices that revolved around each pole.  Depending on the viewing orientation, these vortices could be located to the east of the substellar point, resulting in an apparent westward shift of the latitudinally average temperature structure, the property measured by orbital phase curves.  However, these models also predicted large variations in the shape of the phase curve from orbit to orbit, which has not been seen \citep{Knutson2012}, and strong variability in the depth of secondary eclipses \citep{Rauscher2007a}, at a level ruled out by repeated observations \citep{Agol2010}.

\subsection{Wind and rotation, observed as Doppler shifts in transit spectra}

\begin{figure}[ht!]
\begin{center}
\includegraphics[width=\textwidth]{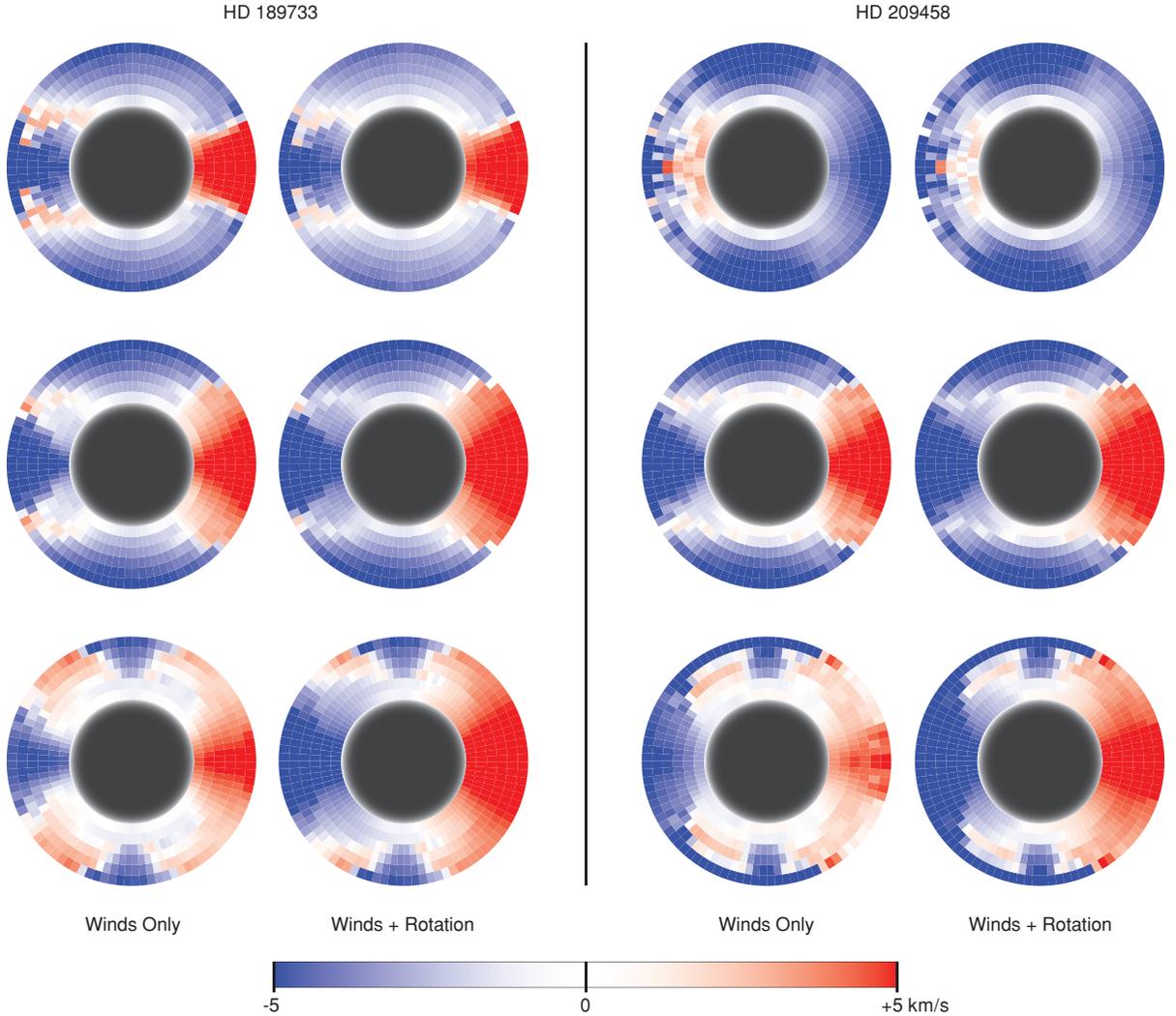}
\end{center}
\caption{Projected line-of-sight (LOS)  velocities across the terminator for each of the 6 models.  The \emph{left} and \emph{right} plots are for models of \HDo~and \HDt, respectively.  The \emph{top, middle}, and \emph{bottom} sets of plots are for $\Prot = 2$, 1, and 0.5 \Porb.  For each pair of images, the one on the left includes the LOS velocity from the winds only, whereas the one on the right includes the effects of both winds and rotation.  For each image, the innermost colored annulus shows LOS velocities at a pressure of 1 bar.  The pressure decreases out to the outer annulus to a pressure of 10 microbars.  The radial scale on these plots has been exaggerated to see detail; in reality the pressure range plotted is less than about 10\% of the planet's radius.  Speeds greater than 5 \kms~in magnitude are shown in the darkest color of red/blue.} \label{fig:termwinds}
\end{figure}

\begin{figure}[ht!]
\begin{center}
\includegraphics[width=0.75\textwidth]{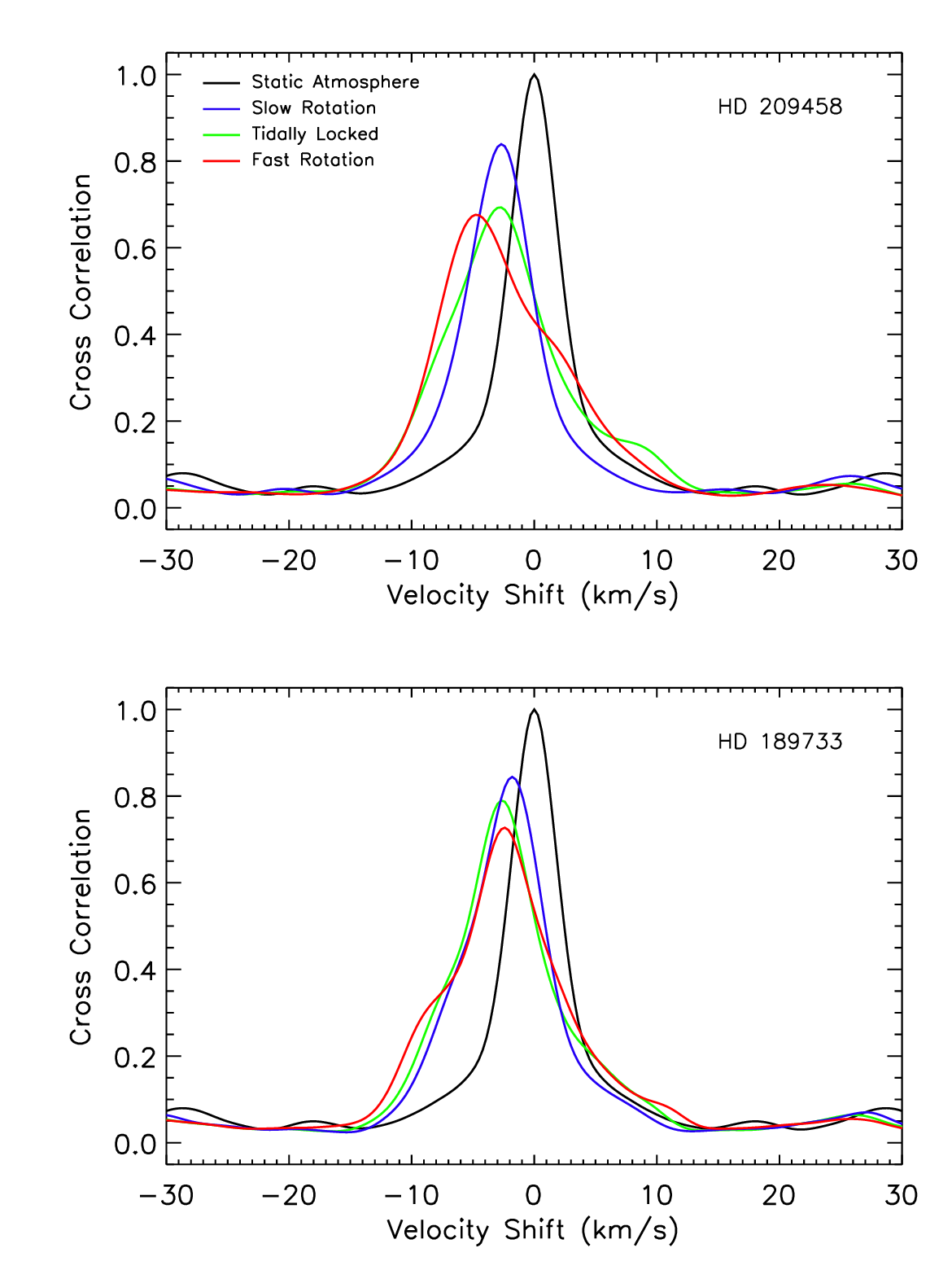}
\end{center}
\caption{Cross correlation functions for transit spectra that include Doppler shifts from both winds and rotation, correlated against an unshifted template spectrum with no winds or rotational broadening included.  For both \HDo~and \HDt , the slowly rotating models produce the smallest net blue shifts.  For reference, the cross correlation function of the template spectrum against itself is shown in black, resulting in a net Doppler shift of 0 \kms.}\label{fig:crscor}
\end{figure}

Figure~\ref{fig:termwinds} shows the wind pattern across the terminator from 1 bar - 10 microbar for all models and helps to explain the net Doppler shift behavior shown in Figure~\ref{fig:crscor} \citep[although the details are more complex than the wind patterns at a single pressure level, see][]{Kempton2012}.  

\begin{figure}[ht!]
\begin{center}
\includegraphics[width=0.7\textwidth, angle=90]{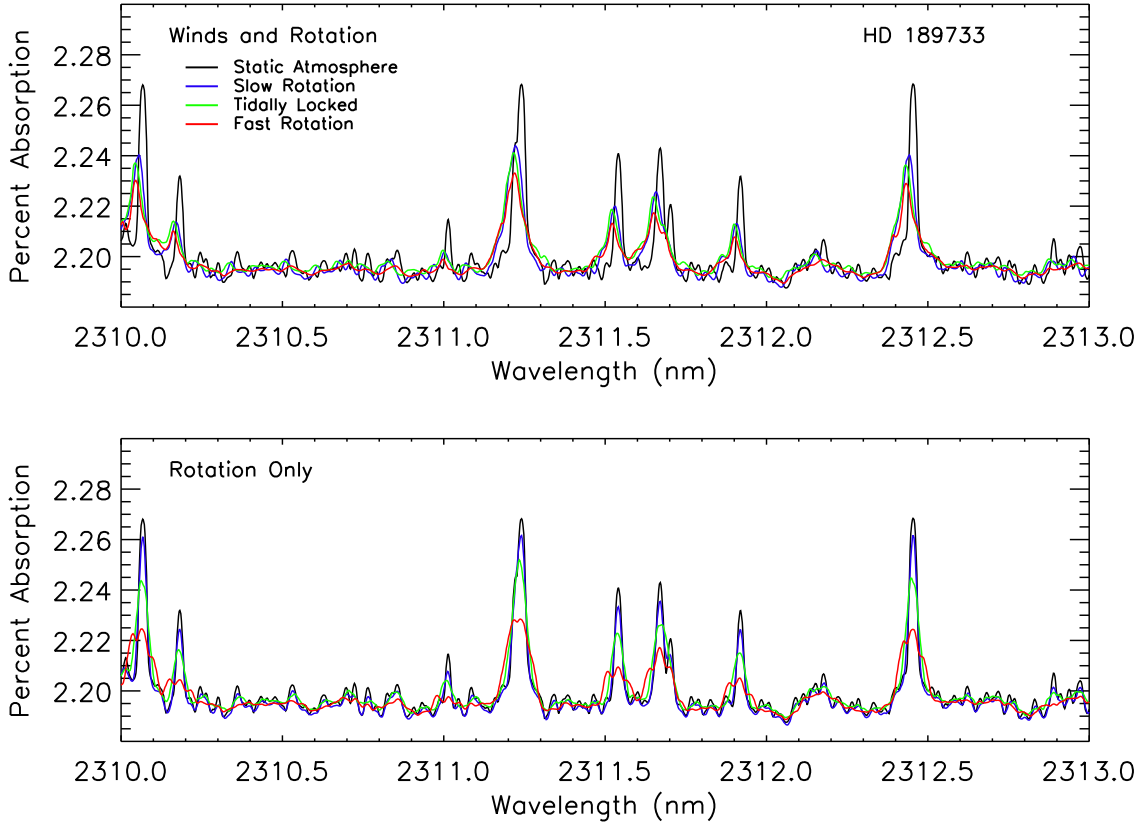}
\end{center}
\caption{Top:  Snapshot of the transit spectrum of HD 189733b over a representative 3-nm wavelength range for slow (blue), locked (green), and fast (red) rotation with Doppler shifts from winds and rotation self-consistently included.  The unshifted transit spectrum obtained for no Doppler shifts is included for reference (black).  A small net blueshift can clearly be seen in all of the Doppler shifted spectra, resulting from strong day-to-night winds in the upper atmosphere of the planet.  Spectral features in this portion of the spectrum mostly result from CO and H$_2$O.  Bottom: Same as above except the Doppler shifts from winds have been removed so as to only show the effects of rotation on the transit spectrum.  Rotational broadening is stronger for faster rotation, resulting in reduced peak line strengths.}\label{fig:HD1spec}
\end{figure}

All of the models of \HDo~(see Figure~\ref{fig:HD1spec}) show strong day-to-night flow across the hot east terminator and slightly weaker flow from night-to-day across the cooler west terminator, in the same sense as the Doppler shift due to rotation.  The peak wind speeds in the slow, locked, and fast cases are similar (11, 9, and 7 \kms~at 0.1 mbar\footnote{The Doppler shifts in the observed spectrum are induced by winds from a range of pressure levels.  Here we quote the maximum winds speeds at the 0.1 mbar level because this roughly the average layer probed by observations \citep{Kempton2012}.}), comparable to the difference in the contribution from the rotational rates ($R_p\omrot =1$, 3, and 5 \kms).  In fact, the net shift, from winds and rotation, are about equal in all three models (see Table~\ref{tab:dopshift}), resulting in very similar net shifts in the cross correlations shown in Figure~\ref{fig:crscor}.  The slight enhancement of blue-shifted components from the more quickly rotating models appears to be due to the slightly hotter eastern terminators of those models, coupled with the faster rotational velocity.  

\begin{figure}[ht!]
\begin{center}
\includegraphics[width=0.7\textwidth, angle=90]{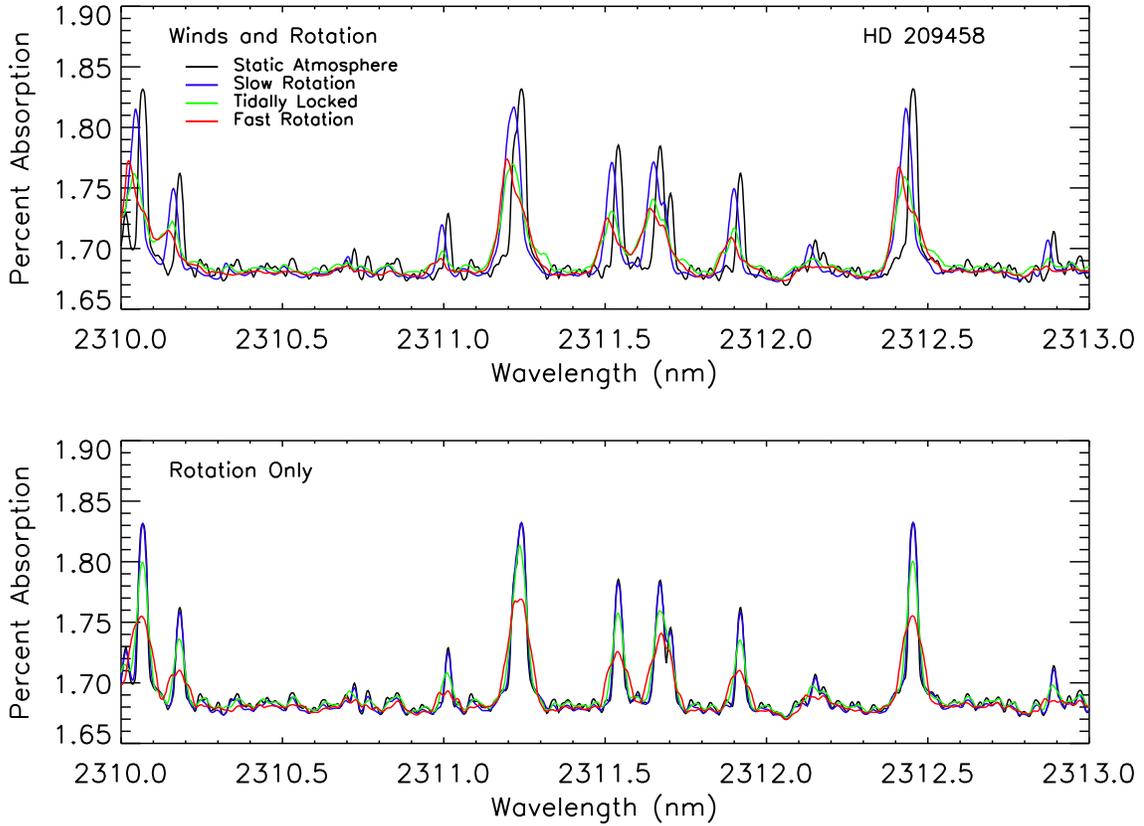}
\end{center}
\caption{Same as Figure~\ref{fig:HD1spec}, but showing spectra instead for HD 209458b.  Asymmetric line profiles for the fast rotation model result from strong day-to-night winds along with the dominant effect of the hot eastern terminator.}\label{fig:HD2spec}
\end{figure}

The models of \HDt~(see Figure~\ref{fig:HD2spec}) are more clearly different in their upper atmosphere wind patterns, as shown in Figure~\ref{fig:termwinds}.  Although the synchronously rotating model has a similar pattern as the synchronous \HDo~model, the fast and slowly rotating models are different.  The fast model has a net day-to-night flow across the terminator (mostly over the poles), and the night-to-day flow across the western terminator is weak.  The slowly rotating model has an even greater difference, with strong day-to-night flow across all regions of the terminator (the night-to-day return flow happens at depth).  For this set of models the Doppler contribution from rotation is $R_p\omrot=1$, 2, and 4 \kms, compared to the maximum wind speeds in the slow, locked, and fast models of 12, 9, and 13 \kms~at 0.1 mbar.  Unifying these trends into a coherent picture, we can understand the net Doppler shift behavior shown in Figure~\ref{fig:crscor} as follows: the blue Doppler shift from the slowly rotating model is dominated by the strong day-to-night winds and minimally affected by the rotation.  In fact, the planet's rotation slightly reduces the net Doppler shift, since the rotation opposes the direction of the winds in the equatorial regions for this model.  The Doppler shift from the tidally locked model is still preferentially blue-shifted, due to stronger day-to-night flow across the eastern terminator, but the significant night-to-day winds across the western terminator also contribute and result in a wider cross correlation peak.  The Doppler shift in the quickly rotating model has a double-peak, albeit uneven, due to the stronger rotational broadening, but the blue-shifted component is dominant, in part because of the very weak night-to-day flow across the western terminator, compared to the stronger day-to-night flow over the poles.  The day-to-night flow across the hot eastern terminator, coupled with the fast rotation speed, further enhances the large net blue-shift obtained for this model.

\begin{deluxetable}{lcc}
\tablewidth{0pt}
\tablecaption{Net Doppler Shifts (\kms)}
\tablehead{
\colhead{}  &  \colhead{HD 189733b} & \colhead{HD 209458b}
}
\startdata
Slow 	    & $-1.8$ 		& $-2.7$ 		\\
Synchronous & $-2.6$		& $-2.8$                \\
Fast        & $-2.4$ 	        & $-4.7$ 	        \\
\enddata
\label{tab:dopshift}
\end{deluxetable}

For both the \HDo~and \HDt~models, we find that the slowly rotating model produces the smallest net blue shift, and there is a weak trend toward larger Doppler shifts as the rotation rate of the planet increases.  This rotation effect can be explained in the following sense.  In all models except for the slowly rotating \HDt, the hotter limb of the planet is the eastern limb.  As a result, the Doppler signature of the atmosphere preferentially comes from the eastern side of the planet because it is physically larger than the cooler western limb \citep[due to a larger scale height, see][]{DobbsDixon2012,Kempton2012}.  In addition to being hotter, the eastern limb typically has strong coherent day-to-night winds, and the rotation of the planet is day-to-night on this side of the planet as well.  These three factors, along with any additional coherent day-to-night flow over the poles and/or western terminator, combine to produce the net blue-shifts obtained for each model.  Because of the asymmetry of the rotational broadening, resulting from the eastern limb of the planet typically being hotter, the faster rotating models tend to produce somewhat larger net Doppler blue shifts.  Only for the slowly rotating \HDt~does the rotation act opposite to the effect of a counter-rotating equatorial flow pattern, and for this model alone, the rotation actually weakens the net Doppler shift.  

As a test of our models, we have separately performed cross correlations of spectra for each model computed with no rotational broadening -- only the Doppler shifts from the winds were included.  For those tests, we found that all three \HDo~models produced net blue shifts of approximately 2 \kms, whereas the \HDt~models produced larger blueshifts of 2.5 - 3 \kms.  This result agrees nicely with the work of \citet{Showman2013}, who showed that weaker Doppler shifts are typically obtained for more weakly irradiated tidally locked hot Jupiters.  In addition to the trend of increasing Doppler shifts for more strongly irradiated planets, our current results now show that planet rotation can further increase or decrease the strength of the Doppler shift, with faster rotating models producing preferentially larger net Doppler shifts.  We also note that our tidally locked \HDt~nicely agrees with the Doppler shift obtained for our tidally locked model without drag from \citet{Kempton2012}.  That result came from a model using the same planetary parameters as our \HDt~model in this paper but employed a Newtonian forcing scheme for the radiative heating, whereas here we are using real radiative transfer.  This is encouraging in that it implies that the results from our previous paper should be fairly reliable, even though the radiative heating was simplified.

\section{Summary and conclusions} \label{sec:conc}

We have presented a study of the observational characteristics from models of hot Jupiters that are not in synchronous rotation states.  We compared models with parameters chosen to represent two well known hot Jupiters, \HDo~and \HDt, the former having a slightly faster synchronous rotation rate than the latter.  For each planet we ran consistent 3D circulation models with rotation rates 0.5, 1, and 2 times its orbital rate ($\Porb=\Prot$ is the standard assumption of a tidally locked, synchronous state).  We found the standard development of an eastward equatorial jet in all models but the slowly rotating model of \HDt, in which the circulation was instead primarily westward.  This indicates the presence of a circulation regime transition at very low rotation rate, although a more detailed characterization and understanding of the regime shift is warranted as a topic for future study.

The observational characteristics we computed from each model were: 1) the thermal emission that would be observed in an orbital phase curve and 2) the net Doppler shift that would be recovered from high-resolution spectra taken during transit.  Neither type of measurement showed much potential as a means by which to constrain the rotation rate of \HDo, as all three models produced very similar signatures.  For \HDt, however, we found that the slowly rotating model would be distinguishable by a delayed peak in the orbital phase curve, while the quickly rotating model would uniquely present a more strongly blue-shifted Doppler signal.  Our results show that in some cases these observational methods could be used to constrain the rotation rate of hot Jupiters, but that further work will be required to determine in more detail the system parameters that enable the use of these techniques.  By modeling a broader swath of parameter space, we can hope to more clearly identify the reasons that our models of \HDt~show observable differences with rotation rate, while our models of \HDo~do not.

We also caution that there are several modeling particularities that could influence the precise values of the Doppler-shift or the phase of peak flux that our models predict.  For example, the strength of numerical dissipation that is applied to a model, necessary to prevent the build-up of small scale noise, is directly related to the wind speeds calculated in the model \citep{Heng2011b} and there are no calculations based on physical conditions that can prescribe what value should be used \citep[see][for some of the complexity involved]{Thrastarson2011}.  Similarly, the temperature structure of the planet and the shape of the orbital phase curve can depend on the particular constituents of the atmosphere.  Specifically, initial data for the planet \HDt~indicated that the atmosphere of this planet may be enhanced in an optical absorber, leading to a temperature inversion in its upper atmosphere \citep[e.g.,][]{Burrows2007b}.  However, recent additional data potentially weaken the evidence for a temperature inversion \citep{Zellem2014}.  Models with stratospheric absorbers have different phase curves and sometimes have different transit spectra properties \citep{Fortney2006,Burrows2010,Fortney2010}, but we do not include any extra absorbers in our models of \HDt.

Finally, in these models we have neglected the magnetic effects that result from the finite thermal ionization in the hot atmospheres and the presence of a planetary magnetic field \citep{Perna2010a,Perna2010b}.  By doing so, we may be missing an important piece of physics, because we have previously shown that these effects can influence atmospheric circulation, particularly for \HDt~(which is hotter and therefore more ionized than \HDo), and alter the observed thermal phase curves \citep{Rauscher2013} and net Doppler shifts \citep{Kempton2012}.  However, the inclusion of magnetic effects in atmospheric circulation models is still in its infancy, with the MHD model by \citet{Batygin2013} indicating that non-zero magnetic fields should preclude day-to-night flow (that all winds should be in the east-west direction), while \citet{Rogers2014} found that the circulation in their (cooler) MHD atmospheres retained the familiar patterns seen in standard GCMs, albeit with slower wind speeds.

The rotation rate of a planet is one of its fundamental characteristics and yet can be very difficult to measure.  We have shown that in some cases in may be possible to combine multiple observational techniques in order to constrain the rotation rate of hot Jupiters.  It would be particularly valuable to measure rotation for this class of exoplanet, since they are typically assumed to have been tidally locked into synchronous rotation and an observational constraint could help to inform theories of tidal dissipation in these systems.  Although more work will be necessary before a set of observations could be uniquely translated into a measured rotation rate, we have demonstrated that this is a fruitful line of theoretical investigation and that the development of more sensitive observations is well motivated.

\acknowledgements

We thank Paul Kempton for designing Figure~\ref{fig:termwinds} and Nick Cowan for providing a thoughtful referee report that helped to improve the quality of this paper.  ER recognizes support from the Sagan Fellowship Program, funded by NASA under contract with the California Institute of Technology (Caltech), and the Lyman P.\ Spitzer Jr.\ Fellowship, awarded by the Department of Astrophysical Sciences at Princeton University.

\bibliography{biblio.bib}

\begin{thebibliography}{80}
\expandafter\ifx\csname natexlab\endcsname\relax\def\natexlab#1{#1}\fi

\bibitem[{{Agol} {et~al.}(2010){Agol}, {Cowan}, {Knutson}, {Deming}, {Steffen},
  {Henry}, \& {Charbonneau}}]{Agol2010}
{Agol}, E., {Cowan}, N.~B., {Knutson}, H.~A., {Deming}, D., {Steffen}, J.~H.,
  {Henry}, G.~W., \& {Charbonneau}, D. 2010, \apj, 721, 1861

\bibitem[{{Arras} \& {Socrates}(2010)}]{Arras2010}
{Arras}, P. \& {Socrates}, A. 2010, \apj, 714, 1

\bibitem[{{Barnes} \& {Fortney}(2003)}]{Barnes2003}
{Barnes}, J.~W. \& {Fortney}, J.~J. 2003, \apj, 588, 545

\bibitem[{{Batygin} {et~al.}(2013){Batygin}, {Stanley}, \&
  {Stevenson}}]{Batygin2013}
{Batygin}, K., {Stanley}, S., \& {Stevenson}, D.~J. 2013, \apj, 776, 53

\bibitem[{{Birkby} {et~al.}(2013){Birkby}, {de Kok}, {Brogi}, {de Mooij},
  {Schwarz}, {Albrecht}, \& {Snellen}}]{Birkby2013}
{Birkby}, J.~L., {de Kok}, R.~J., {Brogi}, M., {de Mooij}, E.~J.~W., {Schwarz},
  H., {Albrecht}, S., \& {Snellen}, I.~A.~G. 2013, \mnras

\bibitem[{{Brogi} {et~al.}(2012){Brogi}, {Snellen}, {de Kok}, {Albrecht},
  {Birkby}, \& {de Mooij}}]{Brogi2012}
{Brogi}, M., {Snellen}, I.~A.~G., {de Kok}, R.~J., {Albrecht}, S., {Birkby},
  J., \& {de Mooij}, E.~J.~W. 2012, \nat, 486, 502

\bibitem[{{Brown}(2001)}]{Brown2001}
{Brown}, T.~M. 2001, \apj, 553, 1006

\bibitem[{{Burrows} {et~al.}(2007){Burrows}, {Hubeny}, {Budaj}, {Knutson}, \&
  {Charbonneau}}]{Burrows2007b}
{Burrows}, A., {Hubeny}, I., {Budaj}, J., {Knutson}, H.~A., \& {Charbonneau},
  D. 2007, \apjl, 668, L171

\bibitem[{{Burrows} {et~al.}(2010){Burrows}, {Rauscher}, {Spiegel}, \&
  {Menou}}]{Burrows2010}
{Burrows}, A., {Rauscher}, E., {Spiegel}, D.~S., \& {Menou}, K. 2010, \apj,
  719, 341

\bibitem[{{Carter} \& {Winn}(2010)}]{Carter2010}
{Carter}, J.~A. \& {Winn}, J.~N. 2010, \apj, 709, 1219

\bibitem[{{Cho} {et~al.}(2003){Cho}, {Menou}, {Hansen}, \& {Seager}}]{Cho2003}
{Cho}, J.~Y.-K., {Menou}, K., {Hansen}, B.~M.~S., \& {Seager}, S. 2003, \apjl,
  587, L117

\bibitem[{{Cho} {et~al.}(2008){Cho}, {Menou}, {Hansen}, \& {Seager}}]{Cho2008}
---. 2008, \apj, 675, 817

\bibitem[{{Cowan} \& {Agol}(2011)}]{Cowan2011a}
{Cowan}, N.~B. \& {Agol}, E. 2011, \apj, 726, 82

\bibitem[{{Cowan} {et~al.}(2007){Cowan}, {Agol}, \& {Charbonneau}}]{Cowan2007}
{Cowan}, N.~B., {Agol}, E., \& {Charbonneau}, D. 2007, \mnras, 379, 641

\bibitem[{{Cowan} {et~al.}(2012){Cowan}, {Machalek}, {Croll}, {Shekhtman},
  {Burrows}, {Deming}, {Greene}, \& {Hora}}]{Cowan2012}
{Cowan}, N.~B., {Machalek}, P., {Croll}, B., {Shekhtman}, L.~M., {Burrows}, A.,
  {Deming}, D., {Greene}, T., \& {Hora}, J.~L. 2012, \apj, 747, 82

\bibitem[{{Crossfield} {et~al.}(2010){Crossfield}, {Hansen}, {Harrington},
  {Cho}, {Deming}, {Menou}, \& {Seager}}]{Crossfield2010}
{Crossfield}, I.~J.~M., {Hansen}, B.~M.~S., {Harrington}, J., {Cho}, J.~Y.-K.,
  {Deming}, D., {Menou}, K., \& {Seager}, S. 2010, \apj, 723, 1436

\bibitem[{{de Kok} {et~al.}(2013{\natexlab{a}}){de Kok}, {Birkby}, {Brogi},
  {Schwarz}, {Albrecht}, {de Mooij}, \& {Snellen}}]{deKok2013b}
{de Kok}, R.~J., {Birkby}, J., {Brogi}, M., {Schwarz}, H., {Albrecht}, S., {de
  Mooij}, E.~J.~W., \& {Snellen}, I.~A.~G. 2013{\natexlab{a}}, ArXiv e-prints

\bibitem[{{de Kok} {et~al.}(2013{\natexlab{b}}){de Kok}, {Brogi}, {Snellen},
  {Birkby}, {Albrecht}, \& {de Mooij}}]{deKok2013a}
{de Kok}, R.~J., {Brogi}, M., {Snellen}, I.~A.~G., {Birkby}, J., {Albrecht},
  S., \& {de Mooij}, E.~J.~W. 2013{\natexlab{b}}, \aap, 554, A82

\bibitem[{{de Wit} {et~al.}(2012){de Wit}, {Gillon}, {Demory}, \&
  {Seager}}]{deWit2012}
{de Wit}, J., {Gillon}, M., {Demory}, B.-O., \& {Seager}, S. 2012, \aap, 548,
  A128

\bibitem[{{Demory} {et~al.}(2013){Demory}, {de Wit}, {Lewis}, {Fortney},
  {Zsom}, {Seager}, {Knutson}, {Heng}, {Madhusudhan}, {Gillon}, {Barclay},
  {Desert}, {Parmentier}, \& {Cowan}}]{Demory2013}
{Demory}, B.-O., {de Wit}, J., {Lewis}, N., {Fortney}, J., {Zsom}, A.,
  {Seager}, S., {Knutson}, H., {Heng}, K., {Madhusudhan}, N., {Gillon}, M.,
  {Barclay}, T., {Desert}, J.-M., {Parmentier}, V., \& {Cowan}, N.~B. 2013,
  \apjl, 776, L25

\bibitem[{{Desch} {et~al.}(1986){Desch}, {Connerney}, \& {Kaiser}}]{Desch1986}
{Desch}, M.~D., {Connerney}, J.~E.~P., \& {Kaiser}, M.~L. 1986, \nat, 322, 42

\bibitem[{{Desch} \& {Kaiser}(1981)}]{Desch1981}
{Desch}, M.~D. \& {Kaiser}, M.~L. 1981, \grl, 8, 253

\bibitem[{{Dobbs-Dixon} \& {Agol}(2013)}]{DobbsDixon2013}
{Dobbs-Dixon}, I. \& {Agol}, E. 2013, \mnras

\bibitem[{{Dobbs-Dixon} {et~al.}(2012){Dobbs-Dixon}, {Agol}, \&
  {Burrows}}]{DobbsDixon2012}
{Dobbs-Dixon}, I., {Agol}, E., \& {Burrows}, A. 2012, \apj, 751, 87

\bibitem[{{Dobbs-Dixon} {et~al.}(2010){Dobbs-Dixon}, {Cumming}, \&
  {Lin}}]{DobbsDixon2010}
{Dobbs-Dixon}, I., {Cumming}, A., \& {Lin}, D.~N.~C. 2010, \apj, 710, 1395

\bibitem[{{Edson} {et~al.}(2011){Edson}, {Lee}, {Bannon}, {Kasting}, \&
  {Pollard}}]{Edson2011}
{Edson}, A., {Lee}, S., {Bannon}, P., {Kasting}, J.~F., \& {Pollard}, D. 2011,
  Icarus, 212, 1

\bibitem[{{Fortney} {et~al.}(2006){Fortney}, {Cooper}, {Showman}, {Marley}, \&
  {Freedman}}]{Fortney2006}
{Fortney}, J.~J., {Cooper}, C.~S., {Showman}, A.~P., {Marley}, M.~S., \&
  {Freedman}, R.~S. 2006, \apj, 652, 746

\bibitem[{{Fortney} {et~al.}(2010){Fortney}, {Shabram}, {Showman}, {Lian},
  {Freedman}, {Marley}, \& {Lewis}}]{Fortney2010}
{Fortney}, J.~J., {Shabram}, M., {Showman}, A.~P., {Lian}, Y., {Freedman},
  R.~S., {Marley}, M.~S., \& {Lewis}, N.~K. 2010, \apj, 709, 1396

\bibitem[{{Gold} \& {Soter}(1969)}]{Gold1969}
{Gold}, T. \& {Soter}, S. 1969, Icarus, 11, 356

\bibitem[{{Guillot}(2010)}]{Guillot2010}
{Guillot}, T. 2010, \aap, 520, A27+

\bibitem[{{Hallinan} {et~al.}(2013){Hallinan}, {Sirothia}, {Antonova},
  {Ishwara-Chandra}, {Bourke}, {Doyle}, {Hartman}, \& {Golden}}]{Hallinan2013}
{Hallinan}, G., {Sirothia}, S.~K., {Antonova}, A., {Ishwara-Chandra}, C.~H.,
  {Bourke}, S., {Doyle}, J.~G., {Hartman}, J., \& {Golden}, A. 2013, \apj, 762,
  34

\bibitem[{{Helled} {et~al.}(2010){Helled}, {Anderson}, \&
  {Schubert}}]{Helled2010}
{Helled}, R., {Anderson}, J.~D., \& {Schubert}, G. 2010, Icarus, 210, 446

\bibitem[{{Heng}(2012)}]{Heng2012}
{Heng}, K. 2012, \apjl, 761, L1

\bibitem[{{Heng} {et~al.}(2011){Heng}, {Menou}, \& {Phillipps}}]{Heng2011b}
{Heng}, K., {Menou}, K., \& {Phillipps}, P.~J. 2011, \mnras, 413, 2380

\bibitem[{{Kataria} {et~al.}(2013){Kataria}, {Showman}, {Lewis}, {Fortney},
  {Marley}, \& {Freedman}}]{Kataria2013}
{Kataria}, T., {Showman}, A.~P., {Lewis}, N.~K., {Fortney}, J.~J., {Marley},
  M.~S., \& {Freedman}, R.~S. 2013, \apj, 767, 76

\bibitem[{{Knutson} {et~al.}(2012){Knutson}, {Lewis}, {Fortney}, {Burrows},
  {Showman}, {Cowan}, {Agol}, {Aigrain}, {Charbonneau}, {Deming}, {D{\'e}sert},
  {Henry}, {Langton}, \& {Laughlin}}]{Knutson2012}
{Knutson}, H.~A., {Lewis}, N., {Fortney}, J.~J., {Burrows}, A., {Showman},
  A.~P., {Cowan}, N.~B., {Agol}, E., {Aigrain}, S., {Charbonneau}, D.,
  {Deming}, D., {D{\'e}sert}, J.-M., {Henry}, G.~W., {Langton}, J., \&
  {Laughlin}, G. 2012, \apj, 754, 22

\bibitem[{{Lecavelier des Etangs} {et~al.}(2013){Lecavelier des Etangs},
  {Sirothia}, {Gopal-Krishna}, \& {Zarka}}]{Lecavelier2013}
{Lecavelier des Etangs}, A., {Sirothia}, S.~K., {Gopal-Krishna}, \& {Zarka}, P.
  2013, \aap, 552, A65

\bibitem[{{Lewis} {et~al.}(2010){Lewis}, {Showman}, {Fortney}, {Marley},
  {Freedman}, \& {Lodders}}]{Lewis2010}
{Lewis}, N.~K., {Showman}, A.~P., {Fortney}, J.~J., {Marley}, M.~S.,
  {Freedman}, R.~S., \& {Lodders}, K. 2010, \apj, 720, 344

\bibitem[{{Li} \& {Goodman}(2010)}]{Li2010}
{Li}, J. \& {Goodman}, J. 2010, \apj, 725, 1146

\bibitem[{{Majeau} {et~al.}(2012){Majeau}, {Agol}, \& {Cowan}}]{Majeau2012}
{Majeau}, C., {Agol}, E., \& {Cowan}, N.~B. 2012, \apjl, 747, L20

\bibitem[{{Mayne} {et~al.}(2014){Mayne}, {Baraffe}, {Acreman}, {Smith},
  {Browning}, {Sk{\aa}lid Amundsen}, {Wood}, {Thuburn}, \&
  {Jackson}}]{Mayne2014}
{Mayne}, N.~J., {Baraffe}, I., {Acreman}, D.~M., {Smith}, C., {Browning},
  M.~K., {Sk{\aa}lid Amundsen}, D., {Wood}, N., {Thuburn}, J., \& {Jackson},
  D.~R. 2014, \aap, 561, A1

\bibitem[{{Menou} {et~al.}(2003){Menou}, {Cho}, {Seager}, \&
  {Hansen}}]{Menou2003}
{Menou}, K., {Cho}, J.~Y.-K., {Seager}, S., \& {Hansen}, B.~M.~S. 2003, \apjl,
  587, L113

\bibitem[{{Merlis} \& {Schneider}(2010)}]{Merlis2010}
{Merlis}, T.~M. \& {Schneider}, T. 2010, Journal of Advances in Modeling Earth
  Systems, 2, 13

\bibitem[{{Miller-Ricci Kempton} \& {Rauscher}(2012)}]{Kempton2012}
{Miller-Ricci Kempton}, E. \& {Rauscher}, E. 2012, \apj, 751, 117

\bibitem[{{Ogilvie} \& {Lin}(2004)}]{Ogilvie2004}
{Ogilvie}, G.~I. \& {Lin}, D.~N.~C. 2004, \apj, 610, 477

\bibitem[{{Pall{\'e}} {et~al.}(2008){Pall{\'e}}, {Ford}, {Seager},
  {Monta{\~n}{\'e}s-Rodr{\'{\i}}guez}, \& {Vazquez}}]{Palle2008}
{Pall{\'e}}, E., {Ford}, E.~B., {Seager}, S.,
  {Monta{\~n}{\'e}s-Rodr{\'{\i}}guez}, P., \& {Vazquez}, M. 2008, \apj, 676,
  1319

\bibitem[{{Perez-Becker} \& {Showman}(2013)}]{PerezBecker2013}
{Perez-Becker}, D. \& {Showman}, A.~P. 2013, \apj, 776, 134

\bibitem[{{Perna} {et~al.}(2012){Perna}, {Heng}, \& {Pont}}]{Perna2012}
{Perna}, R., {Heng}, K., \& {Pont}, F. 2012, \apj, 751, 59

\bibitem[{{Perna} {et~al.}(2010{\natexlab{a}}){Perna}, {Menou}, \&
  {Rauscher}}]{Perna2010a}
{Perna}, R., {Menou}, K., \& {Rauscher}, E. 2010{\natexlab{a}}, \apj, 719, 1421

\bibitem[{{Perna} {et~al.}(2010{\natexlab{b}}){Perna}, {Menou}, \&
  {Rauscher}}]{Perna2010b}
---. 2010{\natexlab{b}}, \apj, 724, 313

\bibitem[{{Polichtchouk} \& {Cho}(2012)}]{Polichtchouk2012}
{Polichtchouk}, I. \& {Cho}, J.~Y.-K. 2012, \mnras, 424, 1307

\bibitem[{{Rasio} {et~al.}(1996){Rasio}, {Tout}, {Lubow}, \&
  {Livio}}]{Rasio1996}
{Rasio}, F.~A., {Tout}, C.~A., {Lubow}, S.~H., \& {Livio}, M. 1996, \apj, 470,
  1187

\bibitem[{{Rauscher} \& {Menou}(2012)}]{Rauscher2012b}
{Rauscher}, E. \& {Menou}, K. 2012, \apj, 750, 96

\bibitem[{{Rauscher} \& {Menou}(2013)}]{Rauscher2013}
---. 2013, \apj, 764, 103

\bibitem[{{Rauscher} {et~al.}(2007){Rauscher}, {Menou}, {Cho}, {Seager}, \&
  {Hansen}}]{Rauscher2007a}
{Rauscher}, E., {Menou}, K., {Cho}, J.~Y.-K., {Seager}, S., \& {Hansen},
  B.~M.~S. 2007, \apjl, 662, L115

\bibitem[{{Rauscher} {et~al.}(2008){Rauscher}, {Menou}, {Cho}, {Seager}, \&
  {Hansen}}]{Rauscher2008}
---. 2008, \apj, 681, 1646

\bibitem[{{Remus} {et~al.}(2012){Remus}, {Mathis}, {Zahn}, \&
  {Lainey}}]{Remus2012}
{Remus}, F., {Mathis}, S., {Zahn}, J.-P., \& {Lainey}, V. 2012, \aap, 541, A165

\bibitem[{{Rodler} {et~al.}(2012){Rodler}, {Lopez-Morales}, \&
  {Ribas}}]{Rodeler2012}
{Rodler}, F., {Lopez-Morales}, M., \& {Ribas}, I. 2012, \apjl, 753, L25

\bibitem[{{Rogers} \& {Showman}(2014)}]{Rogers2014}
{Rogers}, T.~M. \& {Showman}, A.~P. 2014, \apjl, 782, L4

\bibitem[{{Seager} \& {Hui}(2002)}]{Seager2002}
{Seager}, S. \& {Hui}, L. 2002, \apj, 574, 1004

\bibitem[{{Showman} {et~al.}(2010){Showman}, {Cho}, \& {Menou}}]{SCM2010}
{Showman}, A.~P., {Cho}, J.~Y.-K., \& {Menou}, K. {Atmospheric Circulation of
  Exoplanets}, ed. {Seager, S.}, 471--516

\bibitem[{{Showman} {et~al.}(2013){Showman}, {Fortney}, {Lewis}, \&
  {Shabram}}]{Showman2013}
{Showman}, A.~P., {Fortney}, J.~J., {Lewis}, N.~K., \& {Shabram}, M. 2013,
  \apj, 762, 24

\bibitem[{{Showman} {et~al.}(2009){Showman}, {Fortney}, {Lian}, {Marley},
  {Freedman}, {Knutson}, \& {Charbonneau}}]{Showman2009}
{Showman}, A.~P., {Fortney}, J.~J., {Lian}, Y., {Marley}, M.~S., {Freedman},
  R.~S., {Knutson}, H.~A., \& {Charbonneau}, D. 2009, \apj, 699, 564

\bibitem[{{Showman} \& {Guillot}(2002)}]{Showman2002}
{Showman}, A.~P. \& {Guillot}, T. 2002, \aap, 385, 166

\bibitem[{{Showman} \& {Polvani}(2011)}]{Showman2011}
{Showman}, A.~P. \& {Polvani}, L.~M. 2011, \apj, 738, 71

\bibitem[{{Snellen} {et~al.}(2014){Snellen}, {Brandl}, {de Kok}, {Brogi},
  {Birkby}, \& {Schwarz}}]{Snellen2014}
{Snellen}, I.~A.~G., {Brandl}, B.~R., {de Kok}, R.~J., {Brogi}, M., {Birkby},
  J., \& {Schwarz}, H. 2014, \nat, 509, 63

\bibitem[{{Snellen} {et~al.}(2010){Snellen}, {de Kok}, {de Mooij}, \&
  {Albrecht}}]{Snellen2010}
{Snellen}, I.~A.~G., {de Kok}, R.~J., {de Mooij}, E.~J.~W., \& {Albrecht}, S.
  2010, \nat, 465, 1049

\bibitem[{{Snellen} {et~al.}(2009){Snellen}, {de Mooij}, \&
  {Albrecht}}]{Snellen2009}
{Snellen}, I.~A.~G., {de Mooij}, E.~J.~W., \& {Albrecht}, S. 2009, \nat, 459,
  543

\bibitem[{{Socrates} {et~al.}(2012){Socrates}, {Katz}, \&
  {Dong}}]{Socrates2012}
{Socrates}, A., {Katz}, B., \& {Dong}, S. 2012, ArXiv e-prints

\bibitem[{{Spiegel} \& {Burrows}(2013)}]{Spiegel2013}
{Spiegel}, D.~S. \& {Burrows}, A. 2013, \apj, 772, 76

\bibitem[{{Spiegel} {et~al.}(2007){Spiegel}, {Haiman}, \&
  {Gaudi}}]{Spiegel2007}
{Spiegel}, D.~S., {Haiman}, Z., \& {Gaudi}, B.~S. 2007, \apj, 669, 1324

\bibitem[{{Stephens}(1984)}]{Stephens1984}
{Stephens}, G.~L. 1984, Monthly Weather Review, 112, 826

\bibitem[{{Stevenson}(1982)}]{Stevenson1982}
{Stevenson}, D.~J. 1982, Annual Review of Earth and Planetary Sciences, 10, 257

\bibitem[{{Thrastarson} \& {Cho}(2011)}]{Thrastarson2011}
{Thrastarson}, H.~T. \& {Cho}, J.~Y. 2011, \apj, 729, 117

\bibitem[{{Walker} \& {Schneider}(2006)}]{Walker2006}
{Walker}, C.~C. \& {Schneider}, T. 2006, Journal of Atmospheric Sciences, 63,
  3333

\bibitem[{{Warwick} {et~al.}(1989){Warwick}, {Evans}, {Peltzer}, {Peltzer},
  {Romig}, {Sawyer}, {Riddle}, {Schweitzer}, {Desch}, \&
  {Kaiser}}]{Warwick1989}
{Warwick}, J.~W., {Evans}, D.~R., {Peltzer}, G.~R., {Peltzer}, R.~G., {Romig},
  J.~H., {Sawyer}, C.~B., {Riddle}, A.~C., {Schweitzer}, A.~E., {Desch}, M.~D.,
  \& {Kaiser}, M.~L. 1989, Science, 246, 1498

\bibitem[{{Weinberg} {et~al.}(2012){Weinberg}, {Arras}, {Quataert}, \&
  {Burkart}}]{Weinberg2012}
{Weinberg}, N.~N., {Arras}, P., {Quataert}, E., \& {Burkart}, J. 2012, \apj,
  751, 136

\bibitem[{{Welsh} {et~al.}(2010){Welsh}, {Orosz}, {Seager}, {Fortney},
  {Jenkins}, {Rowe}, {Koch}, \& {Borucki}}]{Welsh2010}
{Welsh}, W.~F., {Orosz}, J.~A., {Seager}, S., {Fortney}, J.~J., {Jenkins}, J.,
  {Rowe}, J.~F., {Koch}, D., \& {Borucki}, W.~J. 2010, \apjl, 713, L145

\bibitem[{{Yang} {et~al.}(2013){Yang}, {Cowan}, \& {Abbot}}]{Yang2013}
{Yang}, J., {Cowan}, N.~B., \& {Abbot}, D.~S. 2013, \apjl, 771, L45

\bibitem[{{Zellem} {et~al.}(2014){Zellem}, {Lewis}, {Knutson}, {Griffith},
  {Showman}, {Fortney}, {Cowan}, {Agol}, {Burrows}, {Charbonneau}, {Deming},
  {Laughlin}, \& {Langton}}]{Zellem2014}
{Zellem}, R.~T., {Lewis}, N.~K., {Knutson}, H.~A., {Griffith}, C.~A.,
  {Showman}, A.~P., {Fortney}, J.~J., {Cowan}, N.~B., {Agol}, E., {Burrows},
  A., {Charbonneau}, D., {Deming}, D., {Laughlin}, G., \& {Langton}, J. 2014,
  ArXiv e-prints

\end{thebibliography}

\end{document}